\documentstyle[11pt,epsf,epsfig,myext]{article}
\newdimen\SaveWidth \SaveWidth=\textwidth
\newdimen\SaveHeight \SaveHeight=\textheight
\advance\SaveWidth by -\textwidth
\advance\SaveHeight by -\textheight
\divide\SaveWidth by 2
\divide\SaveHeight by 2
\advance\hoffset by \SaveWidth
\advance\voffset by \SaveHeight

\def\etmiss{\slashchar{E}_T}

\def\ltap{\raisebox{-.4ex}{\rlap{$\sim$}} \raisebox{.4ex}{$<$}}

\let\badcite=\cite
\def\cite{~\badcite}

\def\slashchar#1{\setbox0=\hbox{$#1$}           
   \dimen0=\wd0                                 
   \setbox1=\hbox{/} \dimen1=\wd1               
   \ifdim\dimen0>\dimen1                        
      \rlap{\hbox to \dimen0{\hfil/\hfil}}      
      #1                                        
   \else                                        
      \rlap{\hbox to \dimen1{\hfil$#1$\hfil}}   
      /                                         
   \fi}                                         %

\catcode`@=11
\newdimen\vbigd@men                             

\def\vbig#1#2{{\vbigd@men=#2\divide\vbigd@men by 2%
   \hbox{$\left#1\vbox to \vbigd@men{}\right.\n@space$}}}
\catcode`@=12

\catcode`@=11
\def\citenum#1{\csname b@#1\endcsname}
\catcode`@=12

\begin{document}
\begin{titlepage}
\rightline{LBNL-45198}
\rightline{ATL-PHYS-2000-016}
\begin{center}
{\Large\bf Model Independent Extra-dimension signatures with ATLAS\footnotemark}
\end{center}

\footnotetext{This work was supported in part by the Director, Office of Science,
 Office of High Energy and Nuclear physics, Division of High Energy
Physics of the U.S. Department of Energy under Contract
DE-AC03-76SF00098.}
\bigskip
\centerline{\bf L. Vacavant and I. Hinchliffe }
\centerline{{\it Lawrence Berkeley National Laboratory, Berkeley, CA}}
\bigskip

\begin{abstract}
The generic missing transverse energy signals at LHC  for theories having
large extra dimensions are discussed. Final states of jets plus
missing energy and photons plus missing energy are simulated in the ATLAS
detector. The discovery limit of  LHC and the  methods to 
determine the 
parameters of the underlying model are  discussed.
\bigskip
\end{abstract}
\newpage
\pagestyle{empty}
\tableofcontents
\end{titlepage}

\section{Introduction}

There is much recent theoretical interest in models of
particle physics that have extra-dimensions in addition to the 3+1
dimensions of normal
space-time\cite{Arkani-Hamed:1998},\cite{others}. In these models, new
physics can appear at a mass scale of order 1 TeV and can therefore be 
accessible at LHC.
 The standard model has  a large
hierarchy of scales that exists between the mass scale of the
weak interactions, set by the Fermi constant $G_F$ (or the $W$-mass,
$M_W$), 
and that of gravity,
set by Newton's constant $G_N$ (or the Planck Mass. $M_P \sim 10^{19}$ 
GeV). 
 It is
 believed that a
consistent quantum theory of gravity must be a string theory\cite{string} which
requires additional dimensions in order to be self-consistent. String
theory  has
some inherent scale, the string scale $M_S$, associated with it.
The additional dimensions must be compactified on some scale $R$ so
that they are currently unobserved. It is often  assumed that 
$M_s\sim M_P\sim 1/R$ so that new physics would not be visible until
these huge energy scales are reached. This type of model offers no
insight as to the origin of the large hierarchy although it can, since 
the theory is supersymmetric, ensure that the hierarchy is stable
with respect to quantum corrections. 

Recently however it has been
suggested that $R$ could be much larger, allowing the fundamental scale 
of gravity, here called $M_D$,
 to be close to $M_W$ and so remove the large hierarchy of
scales\cite{Arkani-Hamed:1998}. If there are $\delta$ additional
dimensions of size $R$, then the observed Newton constant is related
to the fundamental scale $M_D$ by
$$G_N=8\pi R^\delta M_D^{-(2+\delta)}$$
If $M_D\sim 1$  TeV then $R\sim 10^{32/\delta -16}$ mm implying that, if 
$\delta \ge 2$, $R$ is smaller than the scales of order 1 mm down to
which gravitational interactions have been probed. In this picture the 
apparent weakness of observed gravity is due to its dilution by the
spreading of its field into the additional dimensions.\footnotemark
\footnotetext{It should be noted that the hierarchy problem is not
  solved in the simplest implementation of the idea; the large ratio
$M_P/M_W$ is replaced by the large value of $RM_D$ whose origin is
not explained.}

When an extra dimension is compactified with  size $R$, say on a
circle, particles propagating exclusively in the extra dimensions
appear,  from a four dimensional viewpoint,  as a tower of massive states. The
characteristic mass splitting of these (Kaluza-Klein) states is of order $1/R$.
 In particular,
gravitons propagating in the extra dimensions will appear to be
massive states whose coupling to ordinary matter is determined only by 
gravitational interactions and is therefore known. However, the
Standard model particles cannot be allowed to propagate into the extra 
dimensions as there is no ``excited electron'' with mass below a 100
GeV.
New physics is expected to appear at a scale $M_D$; the details of
this physics are model dependent.

This paper is concerned primarily with the model independent
signatures and their observability in the ATLAS detector. Some remarks 
about the more model dependent signatures appear in the conclusion.

\section{Graviton direct production}

The emission of gravitons in particle collisions is calculable in terms of
the universal coupling of gravity to all matter ($G_N$). The calculations
become unreliable once energies comparable to the fundamental scale of 
gravity are reached. (Similarly, calculations in the Fermi theory of
weak interactions become unreliable at energy scales of order $M_W$.)
 In the Standard Model, this energy is $M_P$, in
the models with extra dimensions, it is $M_D$. In addition to the
emission of massless gravitons, the Kaluza-Klein excitations that have
mass differences  of order $1/R$ can
also be emitted and their rate calculated as their coupling to
ordinary matter is also
determined.
 For experiments involving
the collision of particles whose energy is much larger than this mass
splitting, the discrete spectrum can be approximated by a continuum
with a density of states $dN/dm \sim m^{\delta -1}$. Since these
emitted gravitons interact very weakly with ordinary matter, their
emission gives rise to missing transverse energy signatures.

 \subsection{Sub-processes and cross-section}

The relevant processes for LHC are $gg\to gG$, $qg\to qG$  and
$q\overline{q}\to Gg$ which give rise to final states of jets plus
missing $E_T$ and $q\overline{q}\to G\gamma$ which gives rise to final
states with a photon plus
missing $E_T$. Final states of $Z$  plus
missing $E_T$ are not considered as the effective rates are much lower 
since the $Z$ can only be observed at LHC via its leptonic decay.

The relevant partonic cross sections
can be written in the form $\frac{d^2\sigma}{dtdm}$ where $m$ is the 
mass of the recoiling (Kaluza-Klein) graviton and $t=(p_a-p_f)^2$ is the usual
Mandelstam variable ($a$ represents and incoming parton and $f$ the
outgoing quark, gluon or photon).

The differential cross-section can be  expressed in the following 
form:
\begin{equation}
\frac{d^{4}\sigma}{dm^{2}dp_{T_{jet,\gamma}}^{2}dy_{jet,\gamma}dy_{G}} =
\frac{m_{G}^{\delta-2}}{2}\frac{S_{\delta-1}}{M_{D}^{\delta+2}}
\frac{d\sigma_{m}}{dt}
\sum_{i,j} \frac{f_{i}(x_{1})}{x_{1}}\frac{f_{j}(x_{2})}{x_{2}}
\label{eq:xsection}
\end{equation}
where the partonic cross-sections $\frac{d\sigma_{m}}{dt}$ 
are given by\cite{Giudice:1999} 
(eqs. 64-67)\footnote{The cancellation of the $\bar{M}_{P}^{2}$ factor 
has already been taken into account in eq.~\ref{eq:xsection}}. $S_{\delta -1}$ is the 
surface of a unit-radius sphere in $\delta$ dimensions and $f_i(x)$
are the parton structure functions.
It is important to notice that the fundamental scale $M_{D}$ is 
factorized in Eqn.~\ref{eq:xsection}: $\sigma \propto M_{D}^{-\delta-2}$.
A change of $\delta$ in this term can be compensated by a change of $M_D$.
Unfortunately, disentangling $M_{D}$ and $\delta$ is difficult. There
is some dependence of $\delta$ induced by the kinematic limit on the
partonic subprocess which implies a limit on the largest value of
the emitted graviton mass $m_G$ that can enter;
 this is discussed in section~\ref{sec:underlying-theory} where the
 variation of the rates with the LHC energy is discussed.
This cross-section is only valid at energies low compared to $M_D$. At 
higher energies, new physics enters and modifies the result. This new
physics is model dependent.

 \subsection{Implementation in ISAJET}

For the purposes of this simulation, the relevant subprocesses have
been implemented in ISAJET\cite{ISAJET} 
and are available in versions 7.48 and later. 
The implementation is modeled on that of $W+jet$, the mass spectrum
being adjusted to reflect the tower of graviton states rather than the 
virtual $W$. 

The process is identified by the name EXTRADIM, and the
keyword
EXTRAD which
expects parameters $\delta$, $M_D$ and a logical flag UVCUT. The last
of these implements a cut off in the cross-section for large values of 
the partonic center of mass energy. If the result depends strongly on
this variable, then the resulting rates are sensitive to physics above 
the scale $M_D$ and the calculations are unreliable. This is 
discussed in more detail in section~\ref{sec:effective-theory}.
The transverse momentum and mass  range of the produced
graviton is specified using the QTW and QMW parameters. Here is an
example:

\vspace{5mm}
\begin{Tabhere}
\begin{minipage}[htb]{0.49\linewidth}
\begin{verbatim}
TEST EXTRADIM G+jet
14000,5000,10,100/
EXTRADIM
QMW
5,1000/
QTW
500,1000/
EXTRAD
2,1000,FALSE/
END
STOP
\end{verbatim}
\end{minipage}
\hspace{2mm}
\begin{minipage}[htb]{0.49\linewidth}
\begin{verbatim}
TEST EXTRADIM G+photon
14000,5000,10,100/
EXTRADIM
QMW
5,1000/
QTW
500,1000/
JETTYPE3
'GM'/
EXTRAD
2,1000,FALSE/
END
STOP
\end{verbatim}
\end{minipage}
\caption{Examples of data cards to generate graviton signals: 
$G+jet$ (left) and $G+\gamma$ (right).}
\end{Tabhere}

 \subsection{Parameter ranges}

\label{sec:range}
 The model has two parameters, the number of extra-dimensions $\delta$ and the 
 fundamental scale $M_{D}$, for which some constraints already exist. 
The reader 
 is referred to Ref~\cite{Giudice:1999} and references therein for further details.
The case $\delta=1$ is already excluded since it would imply 
 deviations of the Newton law of gravitational attraction at distance
 scales that have already been explored. 
 The case $\delta=2$ is not very likely because of cosmological
 arguments. In particular graviton emission from 
 Supernova 1987a \cite{Cullen:1999hc} implies that $M_D>50$ TeV.
 Large values of $\delta$ ($>6$) can not be probed at LHC because
 the cross-section for graviton emission is too small.
 The lower value of $M_D$ that can be considered is determined by two
 factors.
 It should be larger than the current limit from similar processes at
 the Tevatron. Furthermore it should also be large enough so that there
 are no significant contributions from the 
parton-parton 
 center of mass energies where the effective theory is  not  appropriate. 
 This is discussed in section~\ref{sec:effective-theory}.
 
\section{Jet plus missing energy signature}

The signal $pp \rightarrow jet + \etmiss$ is the most promising one for the 
direct graviton production at LHC. The dominant sub-process is 
$qg \rightarrow qG$.
 The following cases have been studied:

 \begin{itemize}
 \item $\delta$ = 2,3,4
 \item $M_{D} = M_{D}^{min},\ldots,5,6,\ldots,10$ TeV
 \end{itemize}
 Event  generation has been performed with ISAJET as explained above.
 The default 
 set of structure function has been used (CTEQ3L).
 Each sample consists of 60000 events generated in $p_{T}^{jet}$ bins and re-weighted.
 The background samples have been generated in the same way and consist each of  300000 events.

 \subsection{Simulation with ATLFAST}

The ATLAS detector was simulated using ATLFAST v2.22\cite{atlfast} with
the default values\footnotemark\footnotetext{It should be noted that to obtain the correct missing 
energy spectrum, the code for the SUSY LSP particle 
in ATLFAST (LPAR6 parameter) has to be set to 39, which is the graviton code.} for low and high-luminosity operation. 
However the transverse energy threshold to accept a  
calorimeter cluster as a jet has been raised from 15 GeV to 30 GeV.
The jet algorithm used is the cone algorithm with a size $\Delta R = 0.4$. 
A larger cone size ($\Delta R = 0.7$)  was also used  but did not lead to 
any improvement. In order to better estimate the direction of the jets, which 
could be useful to fully take advantage of the typically back-to-back topology 
of the signal events, the use of the $k_{T}$ jet algorithm has been investigated 
but no significant improvement was found.

 \subsection{Effective theory and validity range for $M_D$}
 \label{sec:effective-theory}

\begin{figure}
  \[\begin{array}{cc}
  \mbox{\epsfig{file=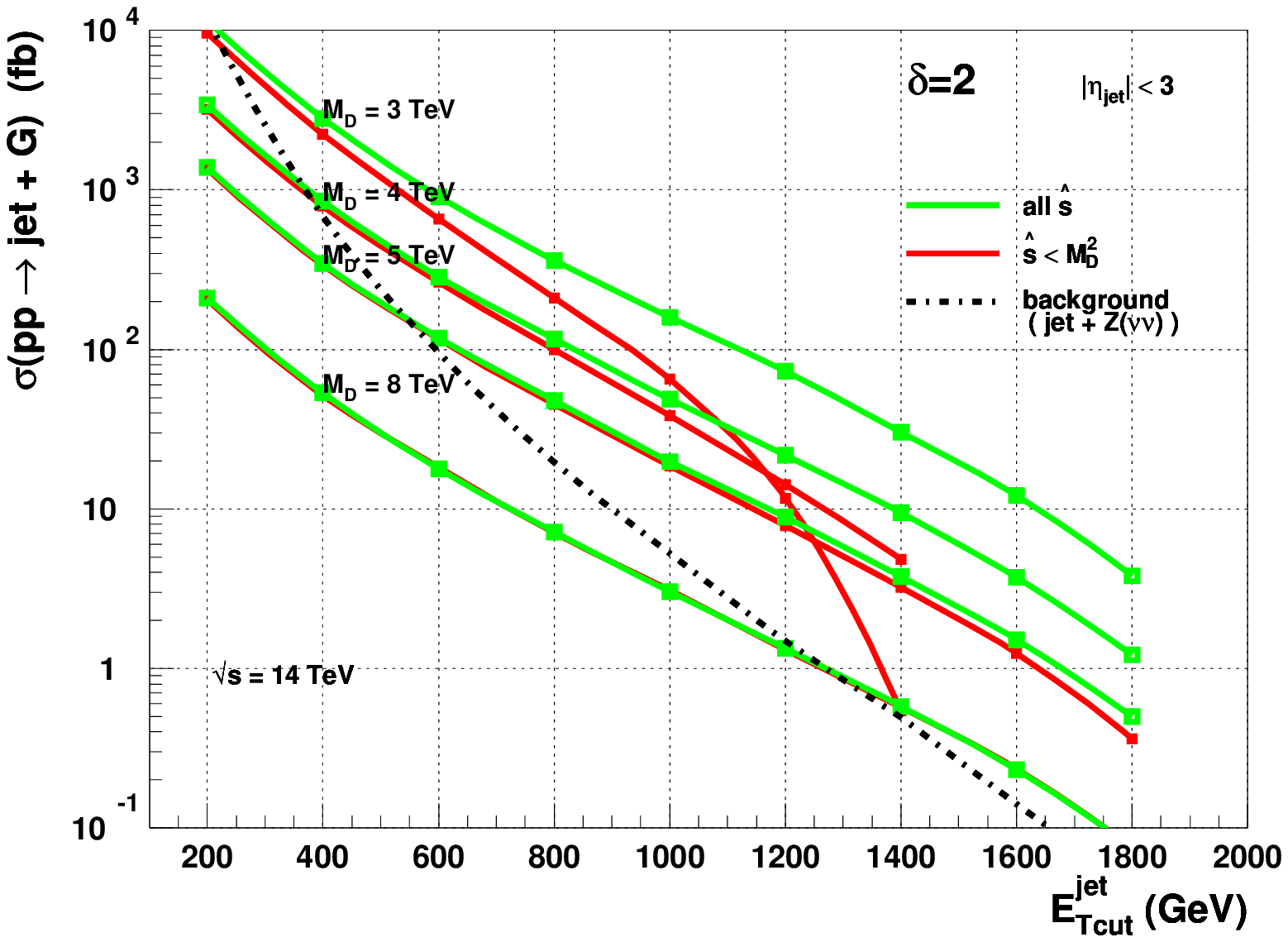,width=0.5\linewidth}}&
  \mbox{\epsfig{file=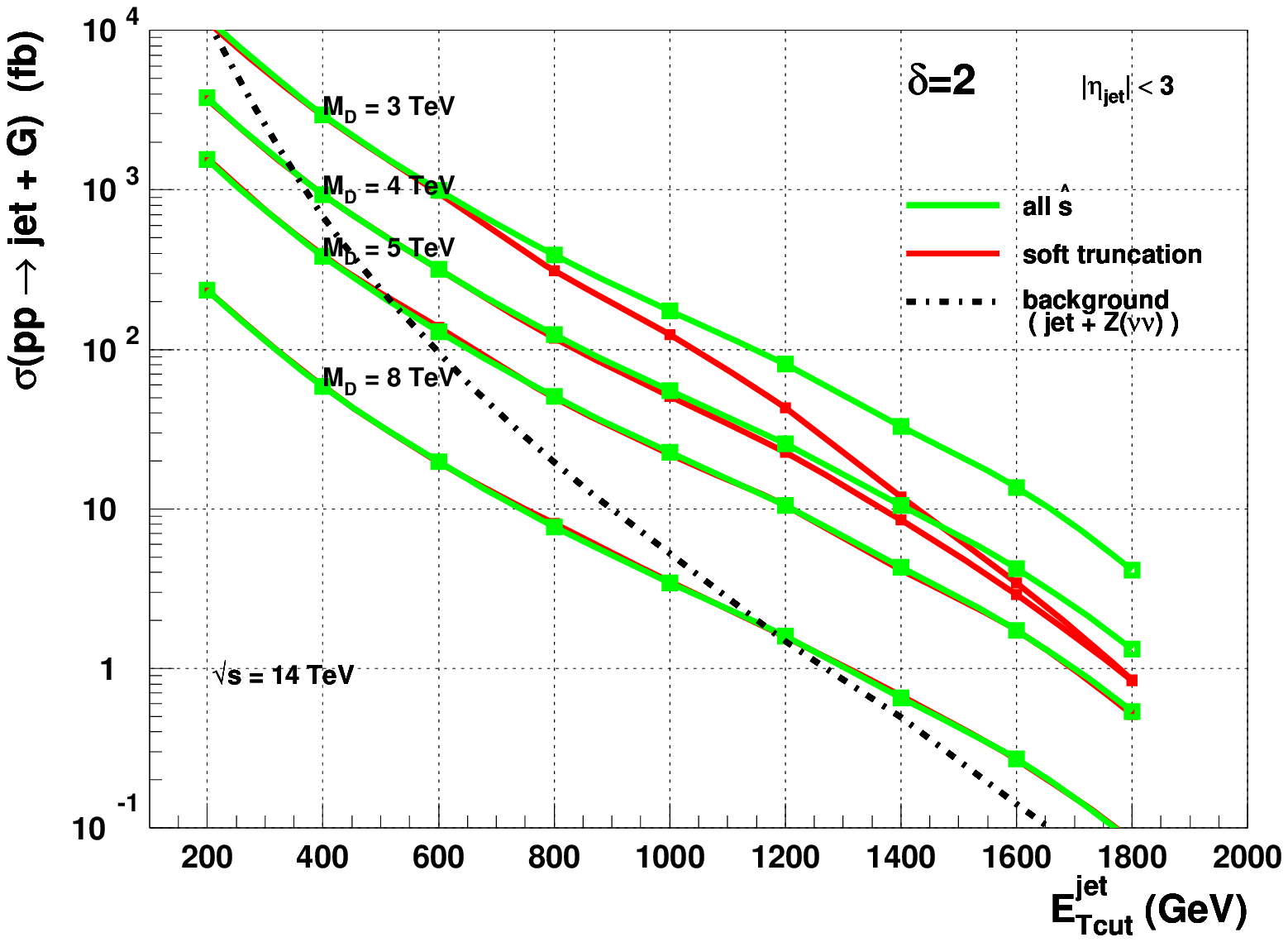,     width=0.5\linewidth}}\\
  \mbox{\epsfig{file=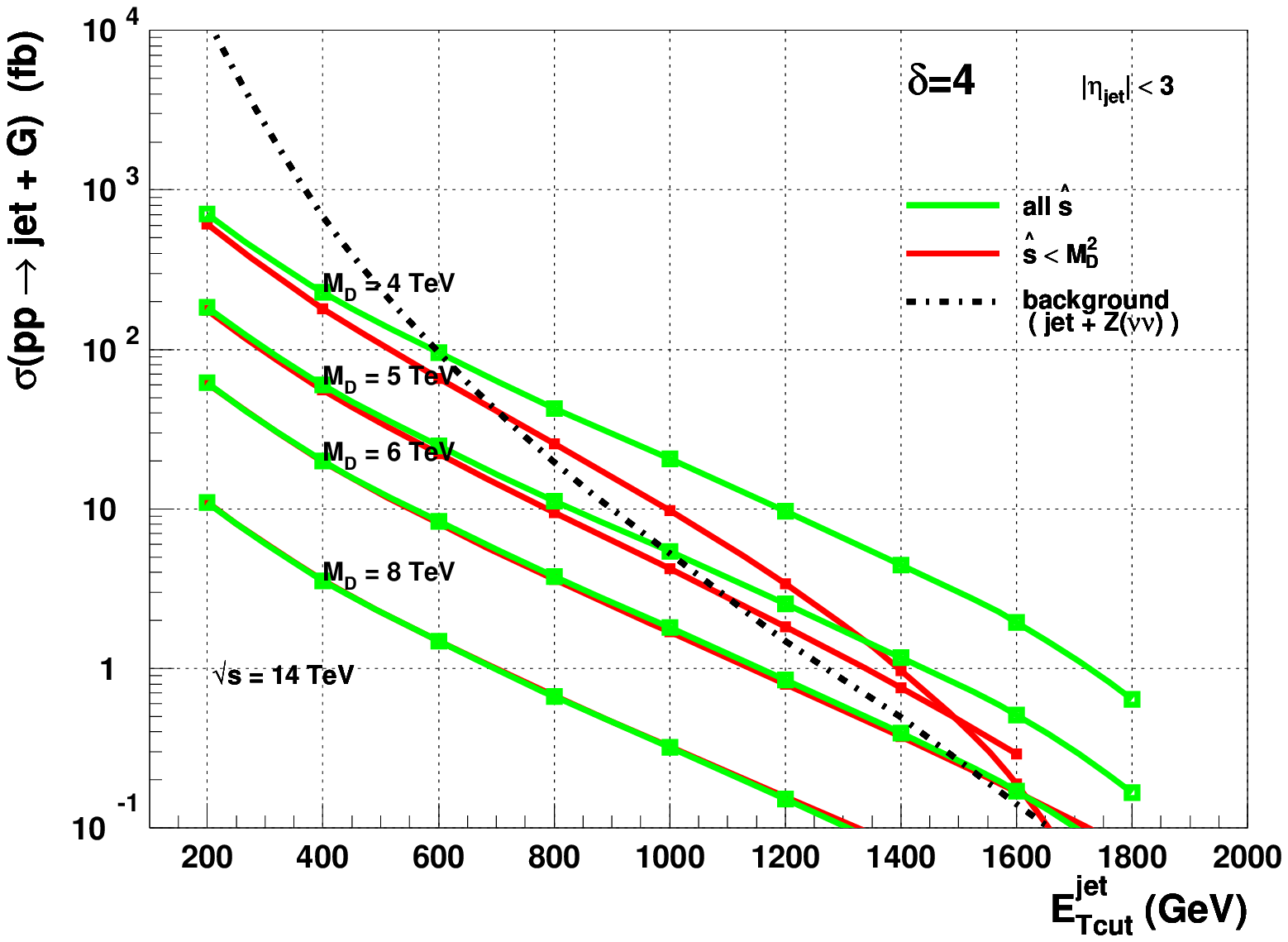,width=0.5\linewidth}}&
  \mbox{\epsfig{file=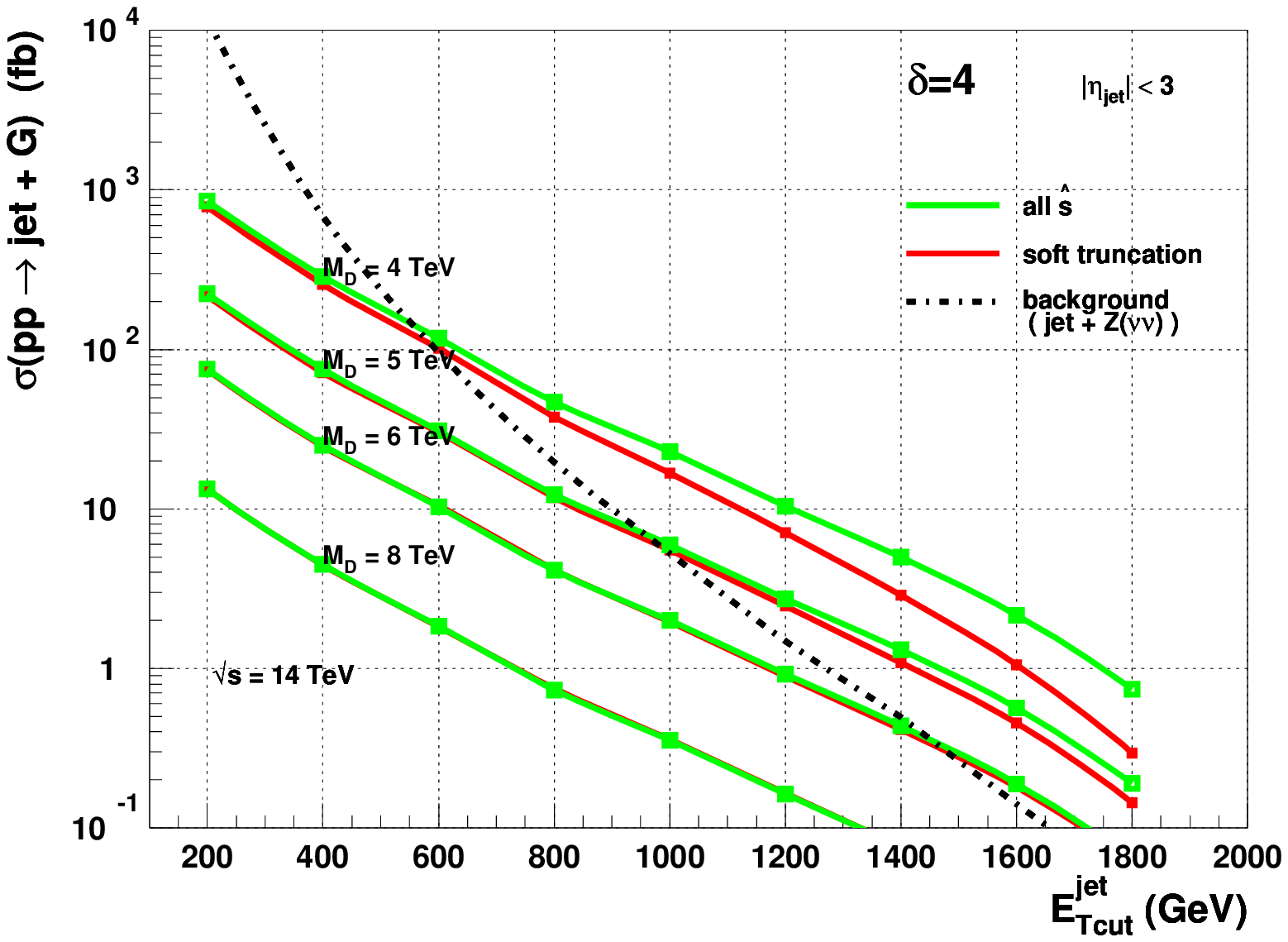,     width=0.5\linewidth}}\\
  \end{array}\]
  \caption{The integrated cross-section $\int_{E_T>E_T^{cut}}
    \frac{d\sigma}{dE_T} dE_T$, 
    for the extra dimensions processes leading to the production of a jet of
    transverse energy $E_T$ in association with missing transverse
    energy at LHC energy. $2$ (top plots) and $4$ extra dimensions are 
 (bottom plots) are 
    used and the curves are labelled by the values of $M_D$. The dashed line 
    shows the main standard model background. The solid black lines show the effect of
    truncating the process according to the prescription of Ref.~\cite{Giudice:1999} 
    (left plots) or following the approach implemented in ISAJET (right plots).}
  \label{fig:cut}
  \end{figure}

 It is important to note that the theory\cite{Giudice:1999} is an 
 effective low-energy theory, valid below the fundamental scale $M_{D}$.
 The behavior above $M_{D}$ is not known, neither is the exact scale at 
 which the theory breaks down.
 To assess the applicability of the effective theory, we 
 checked whether the standard cross-section for
 the process is comparable to the truncated one where contributions from 
 the high energy region are  suppressed. By comparing the
 results obtained by using  Equation \ref{eq:xsection} with those from 
 the truncated form, the region of parameters where the results are
 reliable can be determined; if the results are the same then there is 
 no sensitivity to the high energy region where new physics must
 enter. In Ref.~\cite{Giudice:1999} the partonic cross-section is truncated by
 setting it to zero when
 the partonic center of mass energy ($\sqrt{\hat{s}}$) exceeds $M_D$. In the ISAJET
 implementation a less drastic approach is taken; if the variable
 UVCUT is set then the rate given by Equation  \ref{eq:xsection} is
 reduced by a factor of $M_D^4/\hat{s}^2$ when $\hat{s}>M_D^2$.
The following set of figures illustrate the issues and allow us to
determine the range of parameters in which the theory is applicable. 
Fig. \ref{fig:cut} shows the rates for $\delta=2$ using the cut off
of Ref.~\cite{Giudice:1999} and those using the prescription 
in ISAJET.  The same
quantities for $\delta=4$ are also shown on this figure.

\begin{figure}
\mbox{\epsfig{file=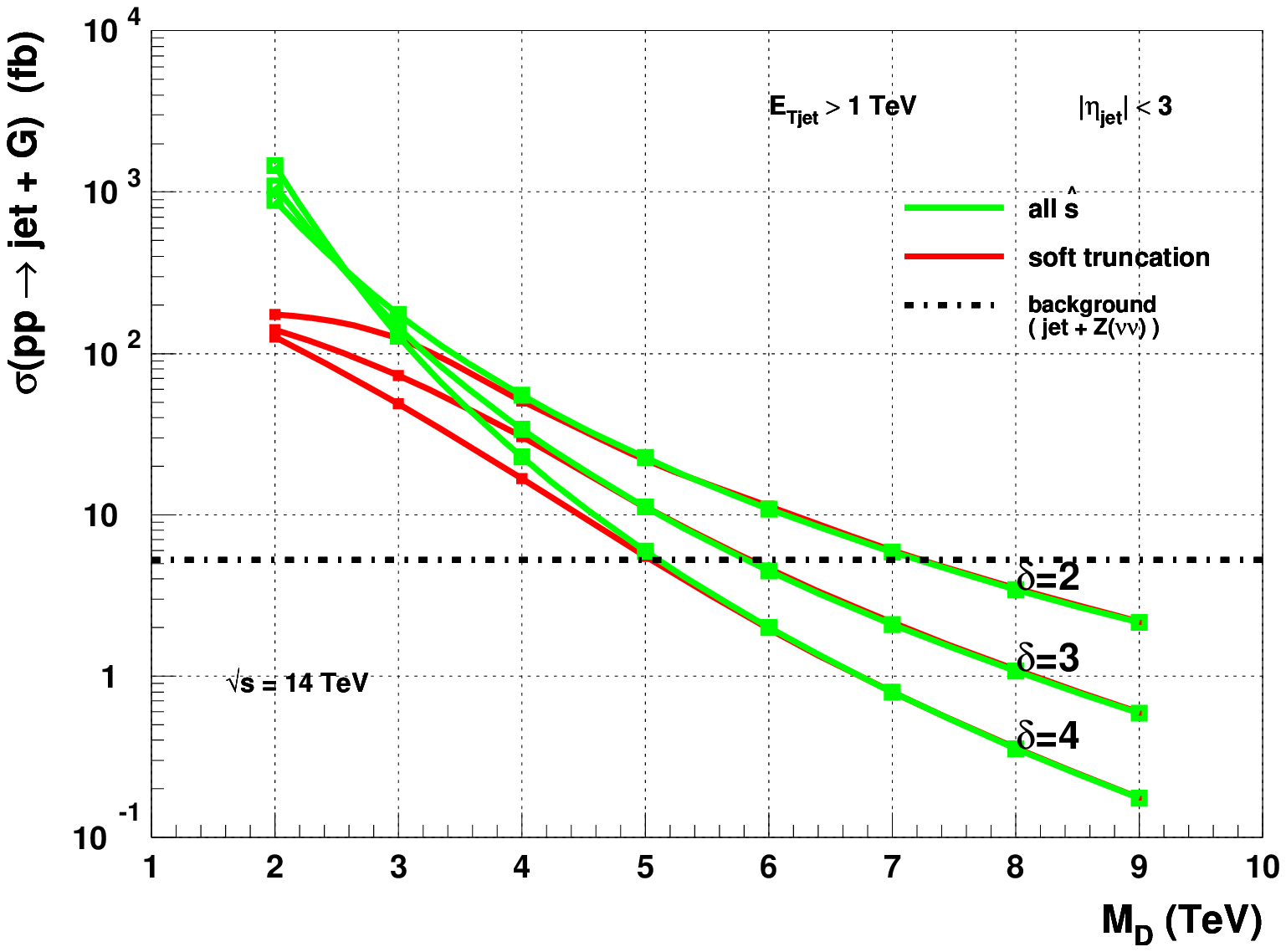,width=\linewidth}}
\caption{Cross section for jets with $E_T>1$ TeV as function of $M_D$
and $\delta$ at 14 TeV. ISAJET is used with UVCUT set to true (dark
lines) and false (light lines). The horizontal line corresponds 
to the standard model background which is approximately
500 events with full luminosity (100 fb$^{-1}$).} 
\label{fig:md1} 
\end{figure}

For fixed values of $M_D$ and $\delta$ there is a maximum value of jet 
$E_T$ for which the results are reliable. Alternatively, for a fixed number of
expected events  or fixed value of $E_T^{cut}$, there is a value of
$M_D$ below which the results are not reliable. If there are
expected to be significant events in this region where the truncated
and untruncated models disagree, then the experiment is sensitive to
new physics appearing at scales $M_D$ and one can expect that other
signals will appear. Fig. \ref{fig:md1} shows the event rate as a
function of $M_D$ and can be used to determine this region. The figure 
shows that the minimum value of $M_D$ increases with $\delta$. For
larger values of $\delta$, $M_D$ must be small in order that the signal 
be visible above the standard model backgrounds. These small values of 
$M_D$ are below the minimum value, hence no model independent
prediction is possible and $\delta\ge 5 $ is not considered here.

 \subsection{Backgrounds}

At the large values of missing $E_T$ that are being considered, the
dominant backgrounds arise from processes that can give rise to
neutrinos in the final state, {\it viz.}  
$jet+Z(\to \nu\nu)$, 
$jet+W(\to \tau \nu)$,
$jet+W(\to \mu \nu)$
and $jet+W(\to e \nu)$. We veto events where there is an isolated
lepton within the acceptance of the ATLAS muon or 
tracking systems as this reduces the background from the last two
sources. An isolated lepton is defined to be one that has less than 10 
GeV of additional energy in a cone of radous $\Delta R=0.2$ around the 
lepton's direction.
Additional instrumental background can arise from events where jet
energies are badly measured or energy is lost in  cracks or
beyond the end of the calorimeter. Studies \cite{etmiss}
have shown that this effect is very small for large values of missing
$E_T$ that are relevant here and we neglect it.

 \subsection{Event Selection and Analysis}
 \label{sec:analysis}

  After the trigger selection, the lepton veto is applied.
  Additional selection relies on the topology of the events: missing 
  transverse momentum, missing transverse momentum and leading jet back-to-back.
  A cut to remove the hadronic decays of the $\tau$ is also described.

If the values of $M_D$ and $\delta$ are such that we are at the limit
of sensitivity, {\it i.e} the signal and the background are comparable 
after missing $E_T$ cuts, one can try to exploit these topological
differences. We will demonstrate that while differences are expected
they are too small to be of significance. 

  \subsubsection{Trigger}

  The trigger is based on a combination of missing 
  energy and jet. At low luminosity, a jet within the trigger acceptance 
  ($|\eta|<3.2$) and with 
  $p_{T} \geq 50$ GeV is required, in addition to at least 50 GeV of missing 
  transverse energy. At high luminosity, both thresholds are raised to 100 GeV.
  If needed, a higher threshold for $\etmiss$ could be used without affecting 
  the study since the extraction of the signal relies on event rates 
  at very large values of
   $\etmiss$ (of the order of one TeV).
  A global efficiency for this selection has been applied, using a 
conservative 
  figure of 90\%. 
  It should be noted that this selection has been performed on the quantities 
  reconstructed by ATLFAST which can slightly differ from the more crudely 
  reconstructed ones at the trigger level.

  \subsubsection{Lepton veto}
  \label{sec:lepveto}

  Events with an isolated lepton are vetoed, mainly to reduce the contribution 
  of the jet+W background where the W decays leptonically. 
  The acceptance for such isolated leptons is defined in ATLFAST as follows:

  \begin{description}
  \item[Electron:]  $p_{T} > 5$ GeV/c, $|\eta|<2.5$
  \item[Muon:] $p_{T} > 6$ GeV/c, $|\eta|<2.5$
  \end{description}

\begin{figure}
  \mbox{\epsfig{file=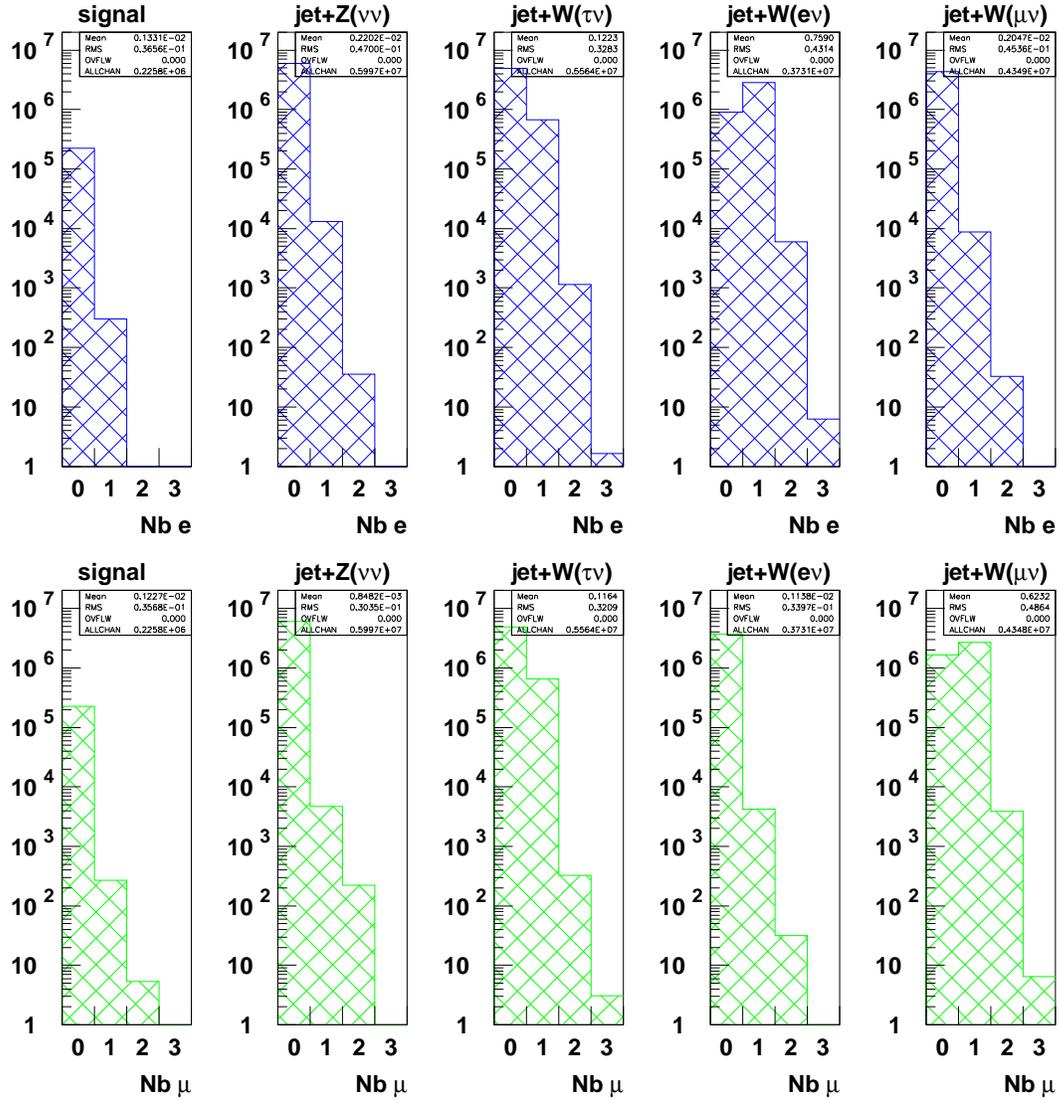,width=\linewidth}}
  \caption{Distribution of the number of isolated electrons and muons
    after trigger cuts for the signal 
  (in this case $\delta=2, \; M_{D}=5$ TeV) and
  for the backgrounds, for 100 fb$^{-1}$.}
  \label{fig:nlept}
  \end{figure}

  The distributions of the number of isolated lepton in the signal and in the 
  backgrounds are shown on Fig.~\ref{fig:nlept}. A conservative value of 
  $\epsilon_{veto}=98\%$ for the lepton veto efficiency has been included by 
  re-weighting the events by $(1-\epsilon_{veto})^{n_{leptons}}$. 
  No provision has been made for difference in the identification 
  efficiency for electrons and muons.
The veto retains 99.8\% of the signal events while rejecting 23.3\%, 74.3\% and 
  61.1\% of the $jW(\tau)$, $jW(e)$ and $jW(\mu)$ background events respectively.

  \subsubsection{Topology}

\begin{figure}\centering
  \mbox{\epsfig{file=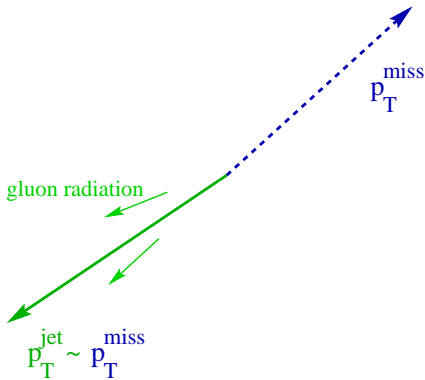,height=5cm}}
  \caption{Sketch of a typical jet+G event.}
  \label{fig:topo}
  \end{figure}

  The topology of the graviton+jet signal is quite simple (Fig.~\ref{fig:topo}): a 
  mono-jet which is back-to-back in azimuth to  balancing missing transverse
  momentum.
Additional jets arise from initial and final state QCD radiation. In
the background events, a $W$ or $Z$ is emitted whereas in the signal
events it is one of the tower of gravitons. The mass of this graviton 
can be large as the density of states increases exponentially with the
mass and a typical emission can therefore have a mass larger than that 
of the $W$. The amount of QCD radiation is controlled by the total energy in the
partonic system. For production at fixed $E_T$, we expect that there
is more energy in the signal process due to the larger mass and
therefore more QCD radiation which should manifest itself in increased 
jet multiplicity. In addition the backgrounds arise from initial
states consisting mainly of $quark-gluon$ collisions, whereas the
signal receives contributions from $gluon-gluon$ initial states. The
greater color charge of the latter again tends to produce more
radiation in the signal events.

  \vspace{5mm}

\begin{figure}
  \mbox{\epsfig{file=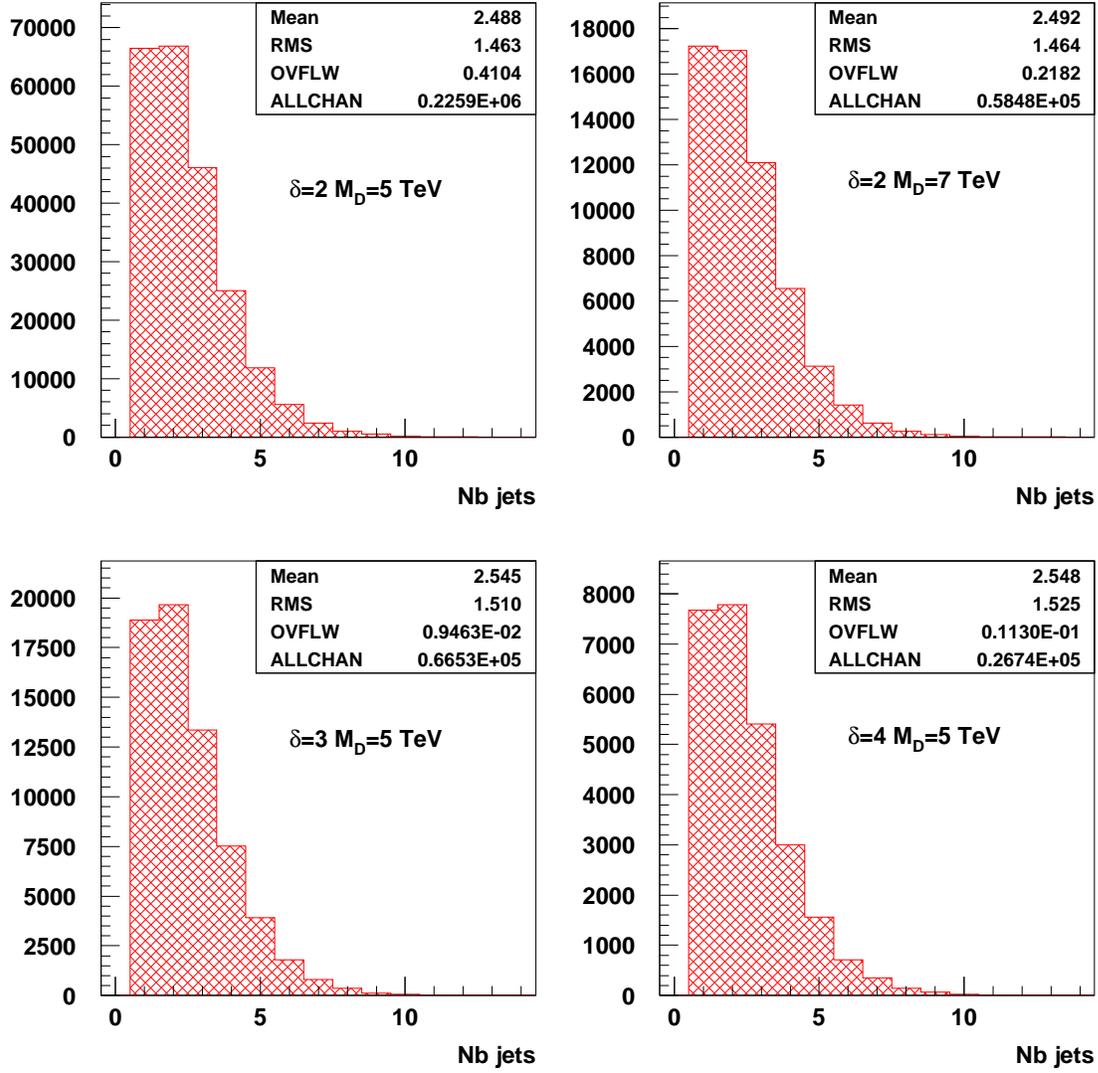,width=\linewidth}}
  \caption{Distribution of the number of jets for different 
  kinds of signal events 
  after the trigger and for one year at high luminosity.}
  \label{fig:njets}
  \end{figure}

\begin{figure}
  \mbox{\epsfig{file=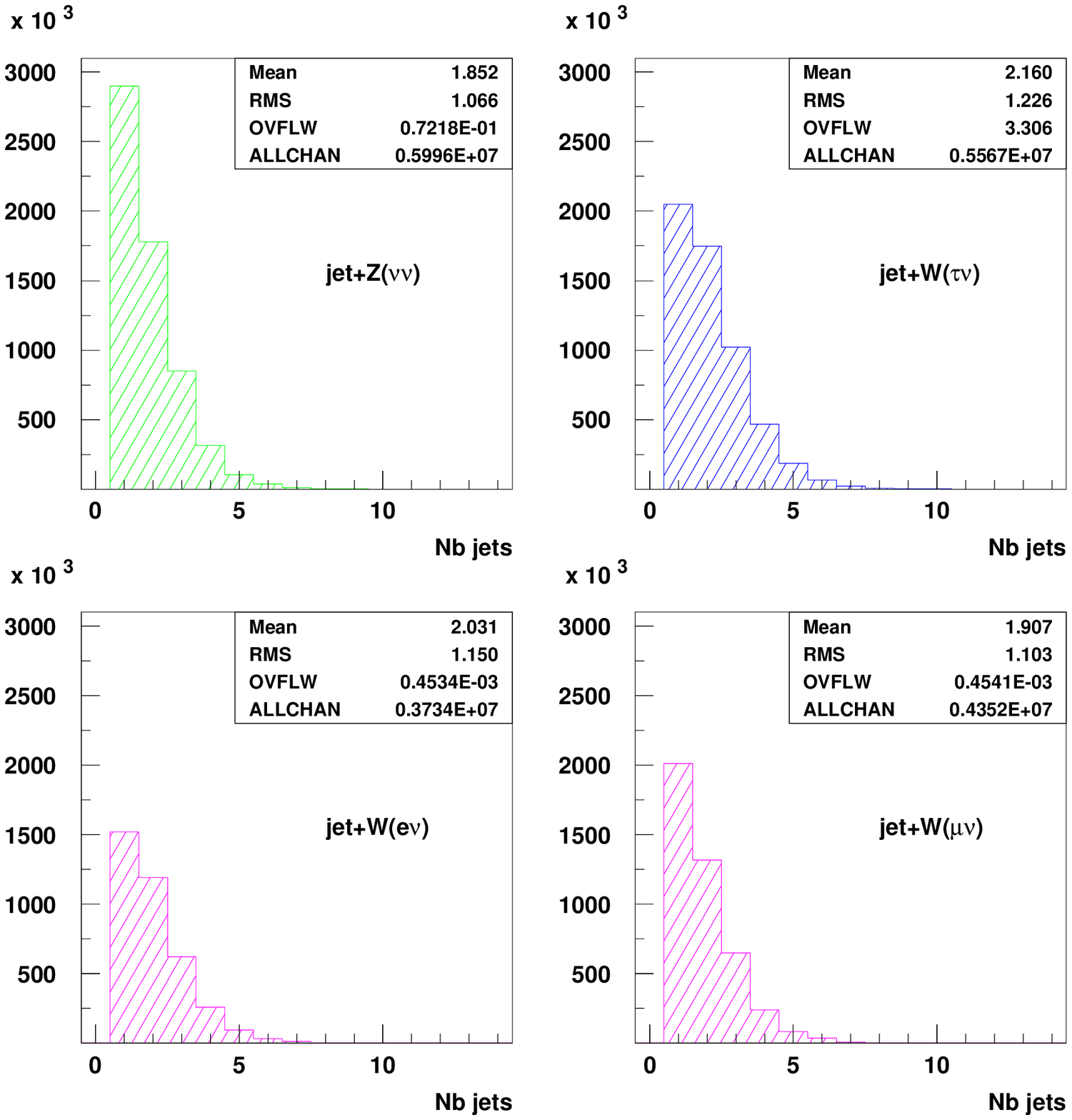,width=\linewidth}}
  \caption{Distribution of the number of jets 
  for the backgrounds after the trigger and for 100 fb$^{-1}$.}
  \label{fig:njetb}
  \end{figure}

  The distribution of the number of jets in the events is shown in 
  Fig.~\ref{fig:njets} for different kinds of signal and in 
  Fig.~\ref{fig:njetb} for the background events. As expected the
  average number of jets in the signal events is larger although the
  difference is small. 
   The average number of jets increases significantly for large 
 missing $E_T$: an average of 4 jets if $\etmiss$ is larger than 
 one TeV. However, the difference between signal and background 
 is very small  for such high values of $\etmiss$.

  \vspace{5mm}

\begin{figure}
  \mbox{\epsfig{file=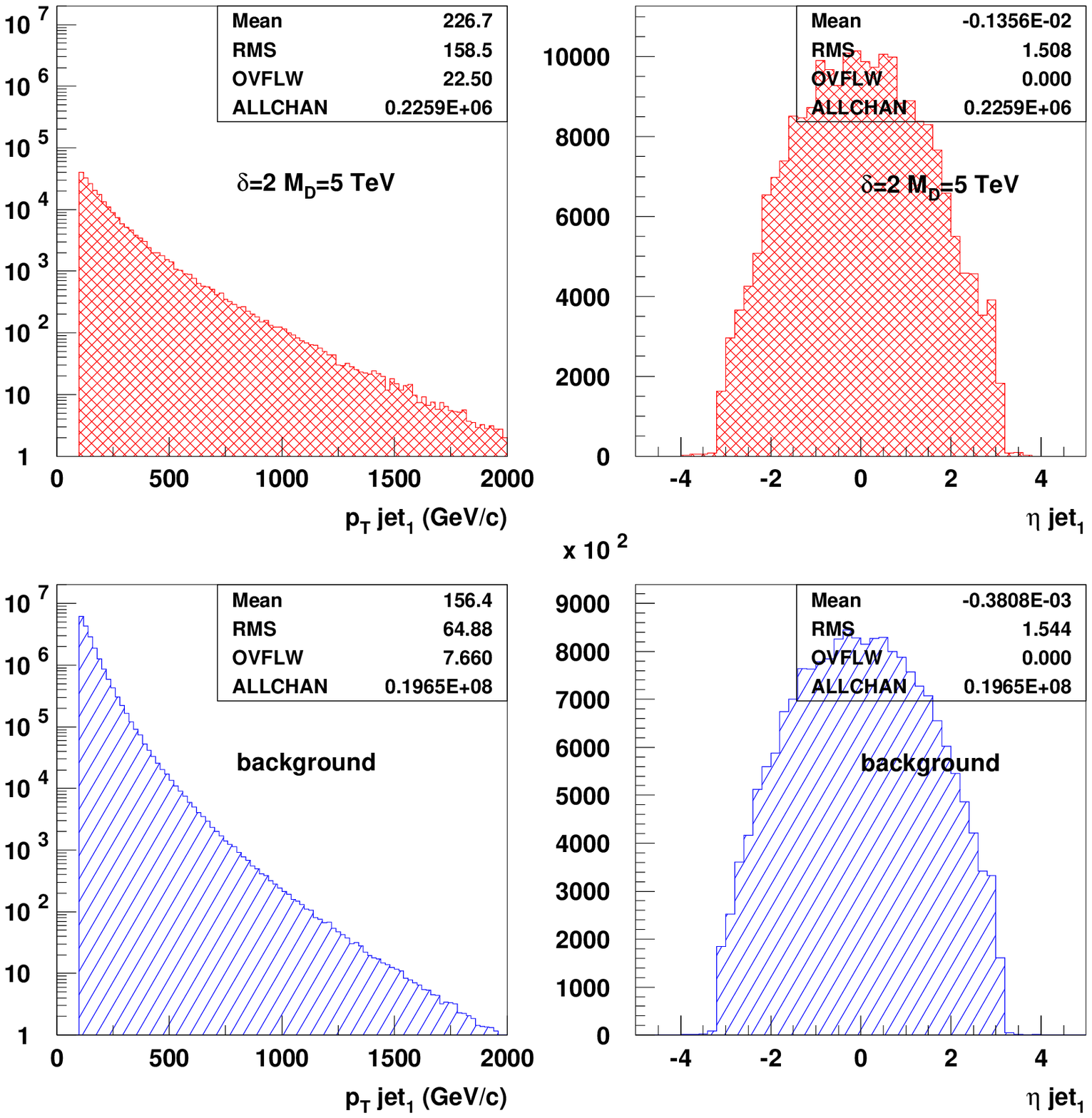,width=\linewidth}}
  \caption{Distributions of the transverse momentum (left) and of the 
  pseudo-rapidity (right) of the leading jet in signal events (top) and for all the 
  backgrounds summed up (bottom),  
  after the trigger and for 100 fb$^{-1}$ of integrated luminosity.}
  \label{fig:jet1}
  \end{figure}

The transverse energy distribution of 
 the jet with highest $E_T$  is shown in Fig.~\ref{fig:jet1}. 
This distribution is strongly correlated with the 
  missing transverse energy in the event. Only the latter will be 
  used for the selection.
 The pseudo-rapidity distribution of the leading jet 
  (Fig.~\ref{fig:jet1}, right) does not 
  allow discrimination between the signal and the backgrounds: in both 
  cases, the jet is rather central.  At lowest order in QCD the missing transverse momentum is back to
  back in azimuth with the leading jet. QCD radiation smears this as can be seen in 
   Fig.~\ref{fig:topo} which  shows the 
  difference between the azimuthal angle of the leading jet and the 
  angle of the missing transverse momentum. The differences between
  signal and background are too small to be useful.

  \vspace{5mm}

  \begin{figure}
  \mbox{\epsfig{file=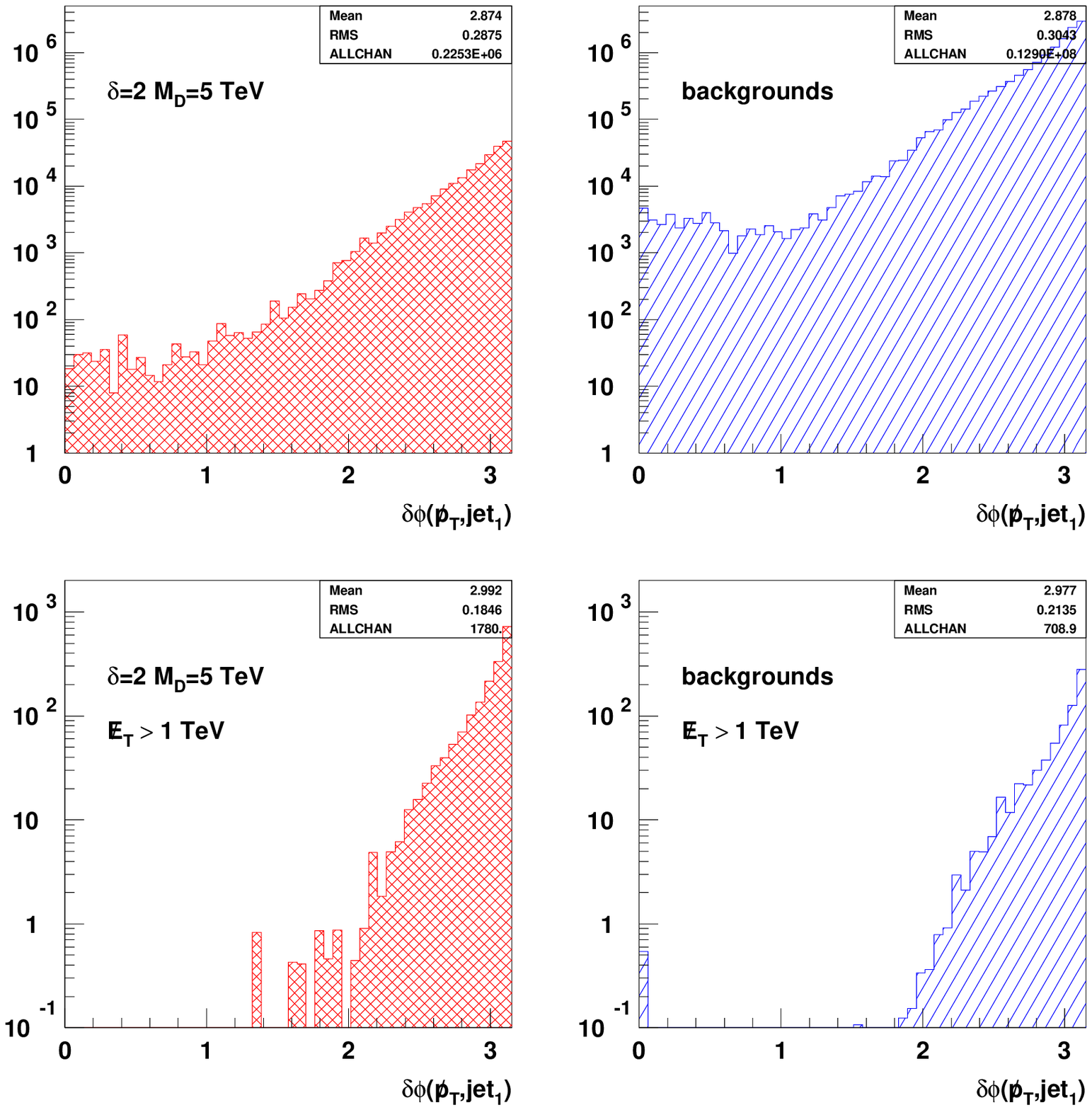,width=\linewidth}}
  \caption{Distributions of the azimuthal angle difference 
  $\delta\phi(\vec{\slashchar{p}}_{T},jet_{1})$ 
  between the missing transverse momentum and the leading jet 
  in signal events (left) and in background events (right) 
  for 100 fb$^{-1}$. The top distributions are after the 
  trigger and the lepton veto while
  the bottom ones include in addition a cut on the missing transverse 
  energy: $\etmiss > 1$ TeV.}
  \label{fig:dfi1}
  \end{figure}
  \vspace{5mm}

\begin{figure}\flushleft
  \mbox{\epsfig{file=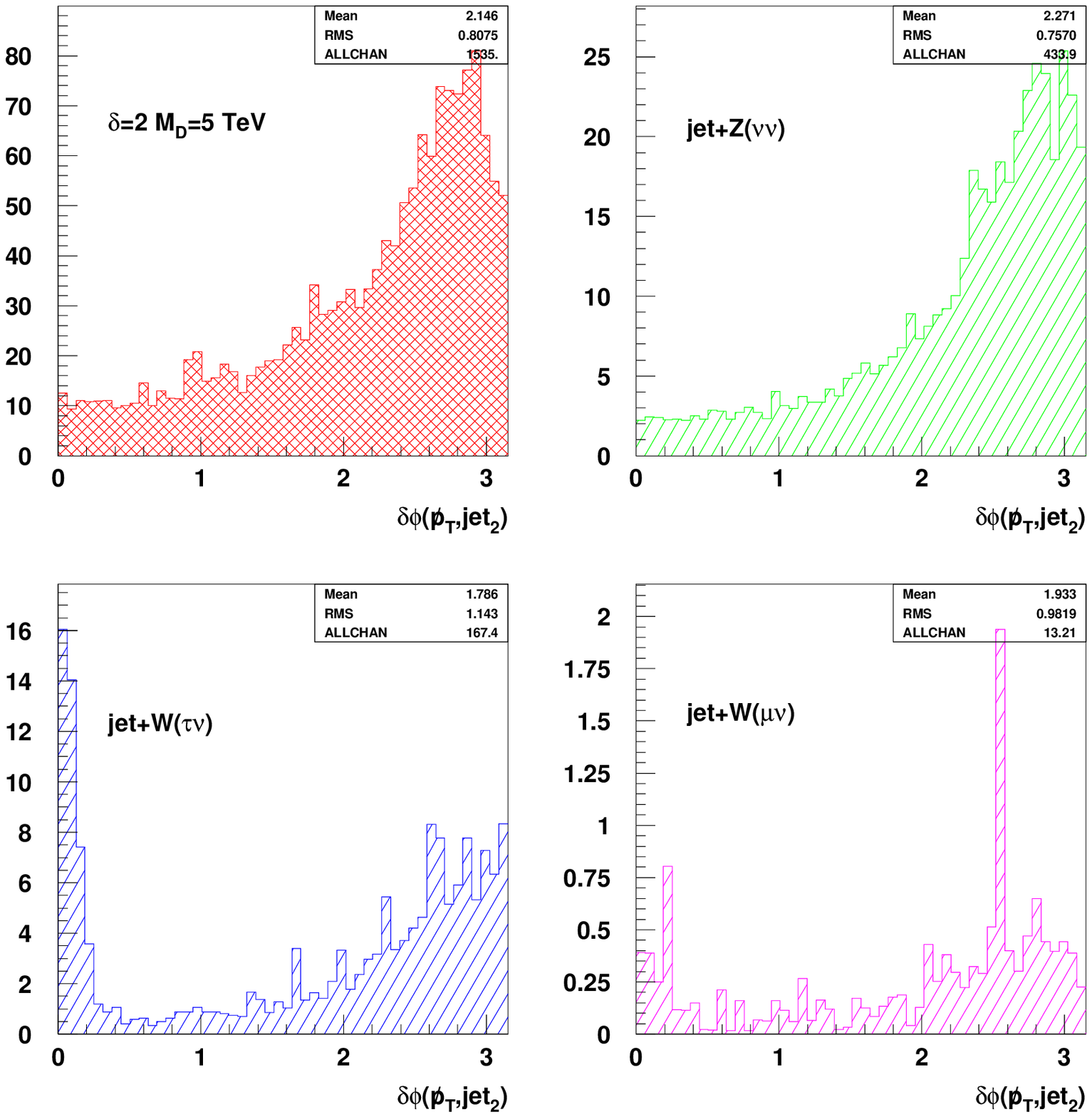,width=\linewidth}}
  \caption{Distributions of the azimuthal angle difference  
  $\delta\phi(\vec{\slashchar{p}}_{T},jet_{2})$ 
  between the missing transverse momentum and the second jet 
  in signal and in background events, 
  after the cut $\etmiss>1$ TeV and for 100 fb$^{-1}$.}
  \label{fig:dfi2b}
  \end{figure}

  In the signal events and in most of the backgrounds, the 
  second jet  is predominantly localized in the opposite 
  hemisphere of the missing transverse energy: the leading jet 
  as well as the other ones induced by gluon radiations and 
  $\vec{\slashchar{p}}_{T}$ are back-to-back in azimuth.
  However this is not the case in the $jet+W(\to\tau\nu)$ events, where 
  hadronic decays of the taus can induce some hadronic activity on the 
  same side of the missing momentum. 
  This feature remains after a large cut on the missing transverse
  energy, as shown on Fig.~\ref{fig:dfi2b} and can be used to
  reduce the background from $W\to\tau\nu$. A requirement that
   $\delta\phi(\vec{\slashchar{p}}_{T},jet_{2}) \geq 0.5$
  rejects 6\% of the signal events, 27\% of the $jW(\tau\nu)$ 
  events and 11\% of all the background.

\begin{figure}
  \mbox{\epsfig{file=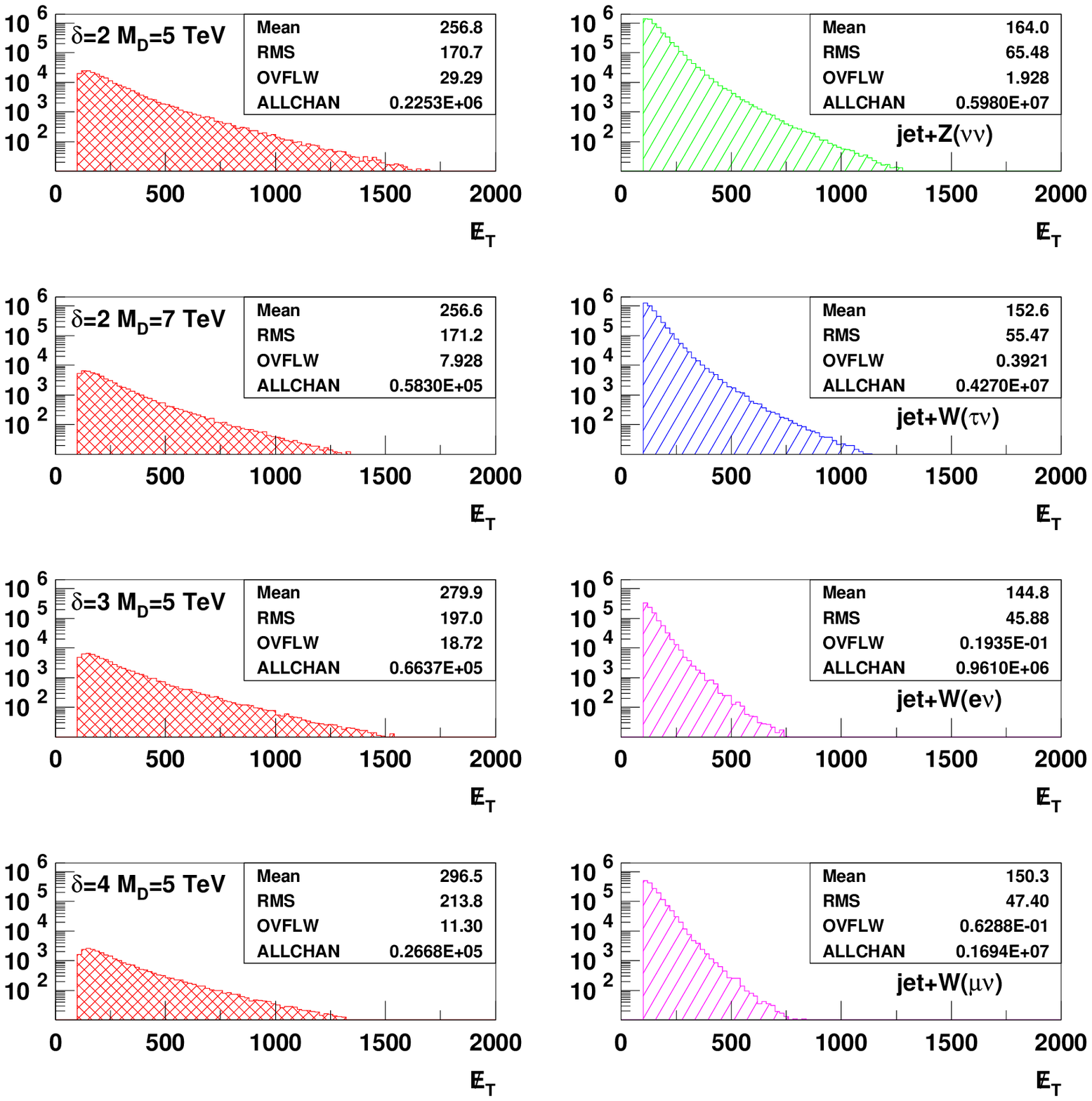,width=\linewidth}}
  \caption{Distributions of the missing transverse energy 
  in signal events (left) and in background events (right)
  after the trigger and the lepton veto and for 100 fb$^{-1}$.}
  \label{fig:etm}
  \end{figure}

\begin{figure}
  \[\begin{array}{cc}
  \mbox{\epsfig{file=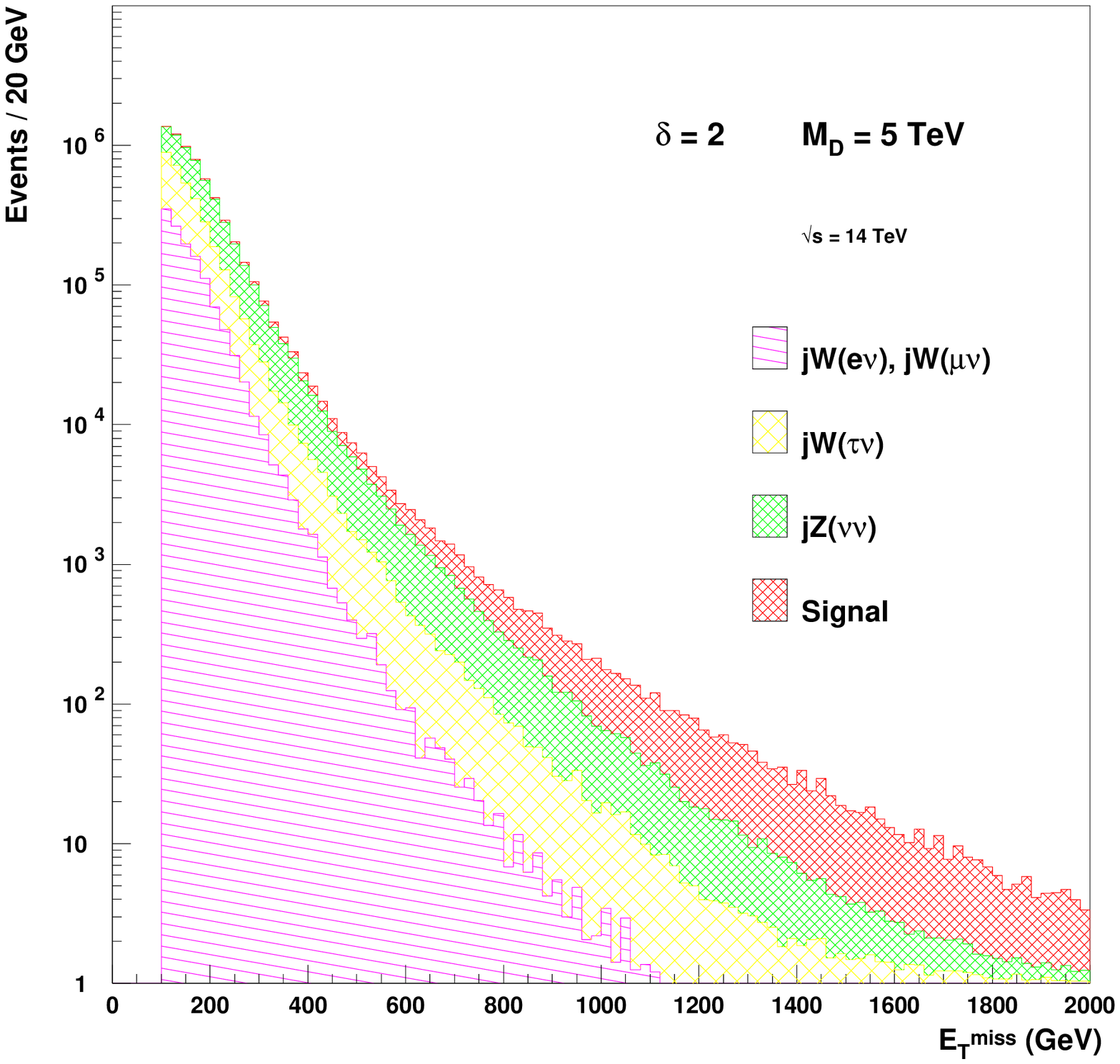,width=0.5\linewidth}}&
  \mbox{\epsfig{file=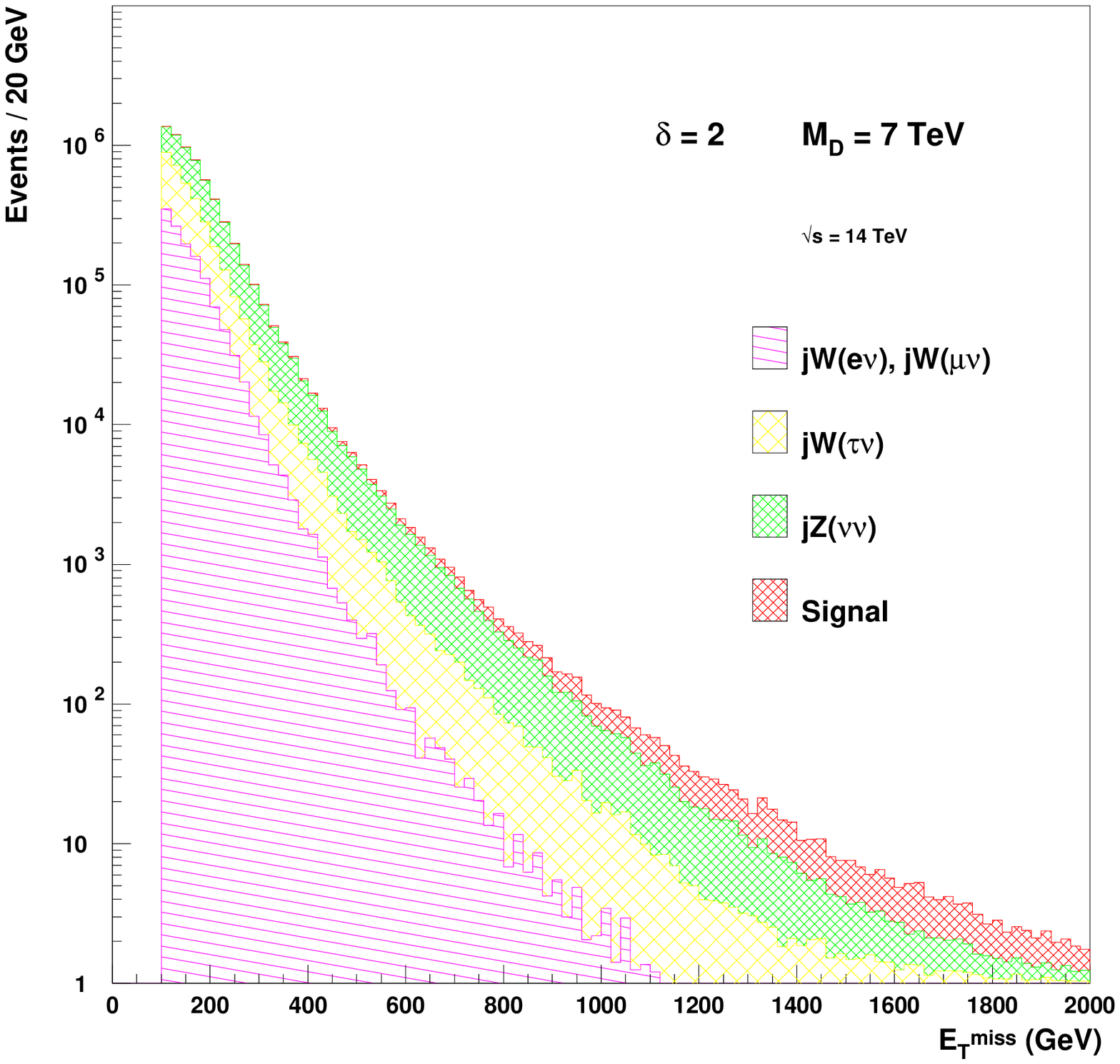,width=0.5\linewidth}}\\
  \mbox{\epsfig{file=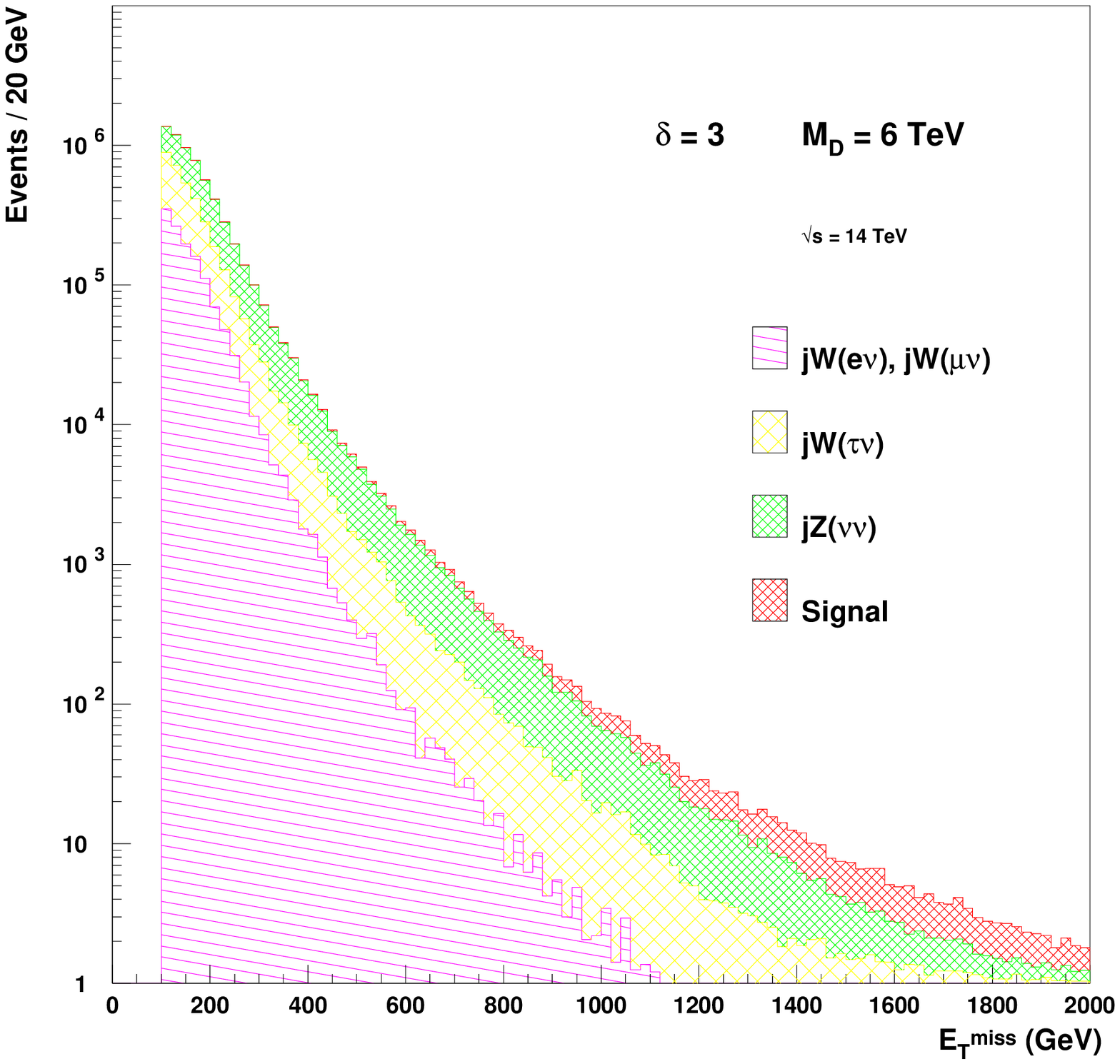,width=0.5\linewidth}}&
  \mbox{\epsfig{file=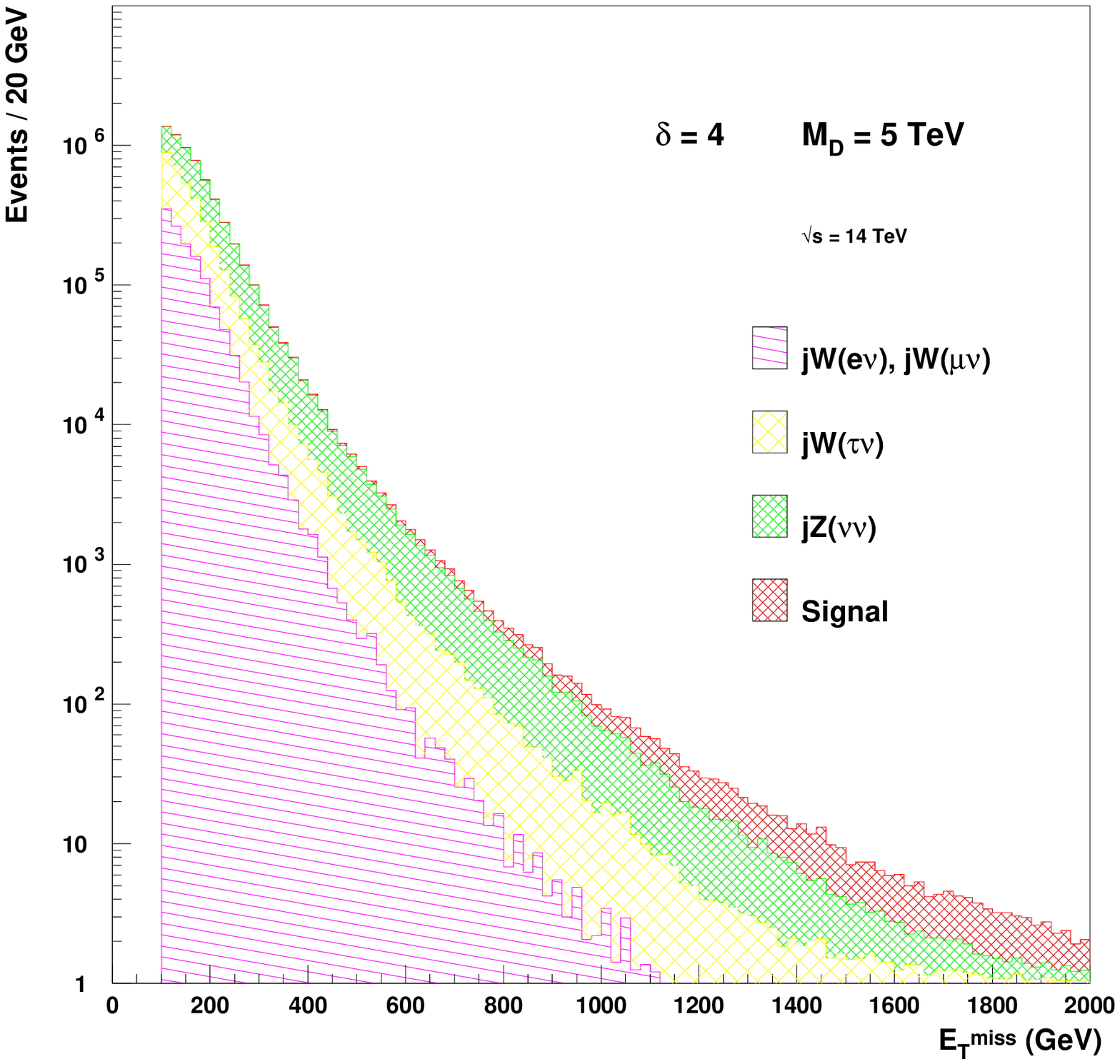,width=0.5\linewidth}}\\
  \end{array}\]
  \caption{Distributions of the missing transverse energy 
  in signal and in background events 
  after the selection and for 100 fb$^{-1}$ of integrated luminosity. Various 
  cases $(\delta,M_{D}$) for the signal are shown.}
  \label{fig:etm1}
  \end{figure}

Finally,  Fig.~\ref{fig:etm} shows the missing transverse energy
  distribution of the signal and backgrounds. The dominant background
  arises from the 
  $jZ(\nu\nu)$ 
final state; $jW(\tau\nu)$ being the next most 
significant.
 Fig.~\ref{fig:etm1} shows the missing transverse momentum
  distributions for several choices of $\delta$ and $M_D$; the signal
  can be seen to emerge from the background at large
  $E_T^{miss}$. These show the expected scaling 
  of the cross-section as a function of $M_{D}^{-\delta-2}$.

 \subsection{Sensitivity}

In this section we estimate the sensitivity of ATLAS to $\delta$ and
$M_D$.
  The background is dominated by the $jZ(\nu\nu)$ events which can be 
calibrated 
  using events with electronic and muonic decays of the $Z$.
  Since $Br(Z\rightarrow \nu\nu) \sim 3[ Br(Z\rightarrow\mu\mu) +
  Br(Z\rightarrow ee)]$ and the acceptance for muons and electrons
  from $Z$ is not complete the size of this sample is smaller than the 
  background and hence, if it is used to normalize the background, the
  error is larger than the naive estimate of the square root of the
  number of expected background events.

The ratio between the background and the calibration samples has been 
studied by simulating and applying a basic selection on 
$jZ(\rightarrow ee)$ events. For triggering we required 
one jet of at least 50 (100) GeV at low (high) luminosity within the 
trigger acceptance ($|\eta| < 3.2$) and two isolated electrons of 
at least 15 (20) GeV within $|\eta|<2.5$. The invariant mass of the two 
electrons is required to lie in $m_{Z}\pm 10$ GeV.

\begin{figure}
\mbox{\epsfig{file=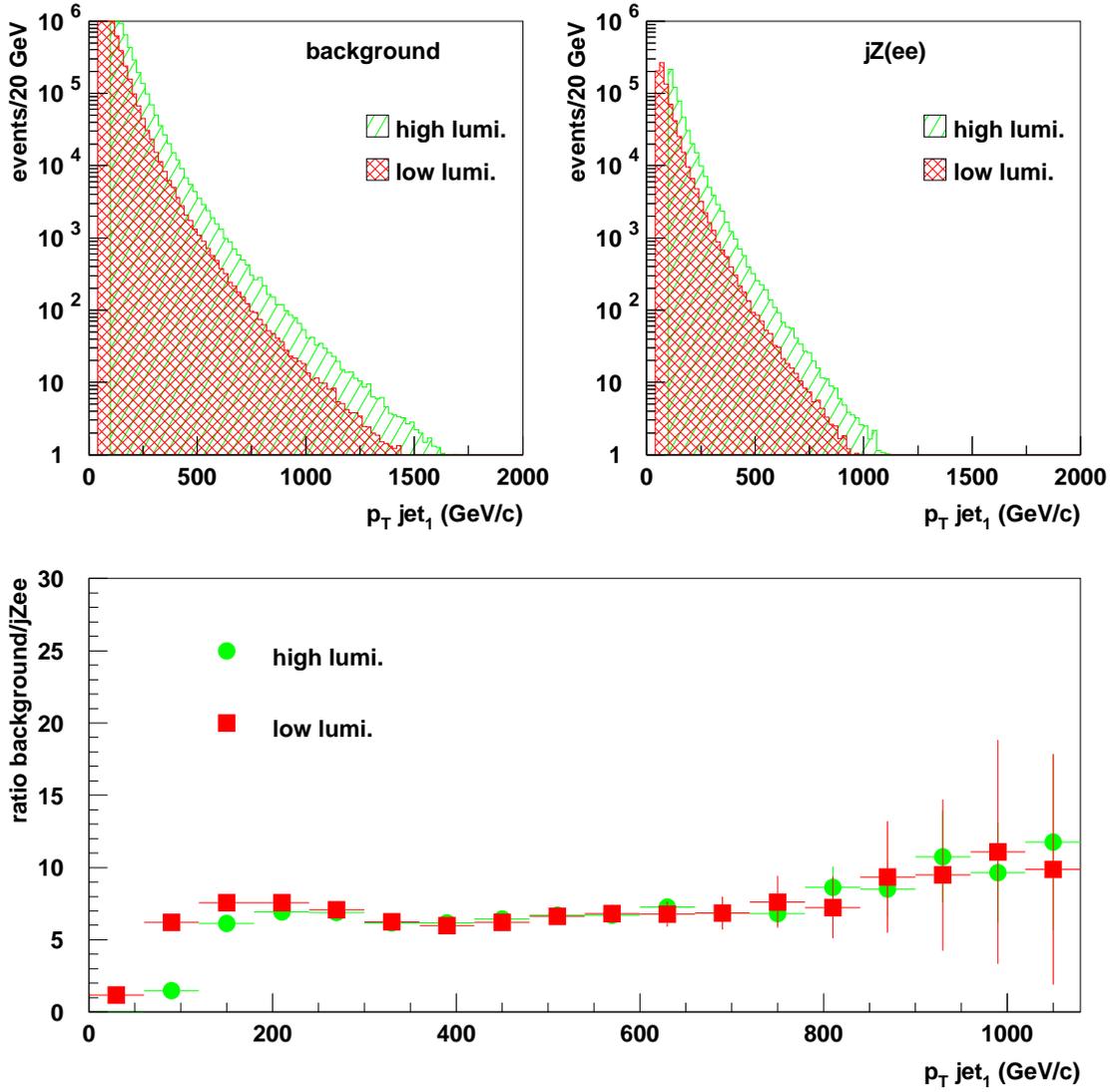,width=\linewidth}}
\caption{Distribution of the transverse momentum of the 
leading jet in background events (top left) and in the 
$jZ(\rightarrow ee)$ calibration sample (top right). The 
bottom plot shows their ratio (an extra factor of 0.5 has been 
included in the ratio to take into account the 
$jZ(\rightarrow \mu\mu)$ calibration sample).}
\label{fig:norma}
\end{figure}

Assuming that the $jZ(\to\mu\mu)$ sample can also be used, the
calibration sample can be doubled. It can be seen from Fig.~\ref{fig:norma} 
that the calibration sample is approximately a  factor of 7 smaller
than the  background sample. We therefore used
   $S/\sqrt{7B}$ as a 
statistical estimator of the signal significance. This is a worst case 
scenario; in practice the many measurements at LHC will give one
confidence in predicting the missing transverse energy rate in the
signal region.

Table~\ref{tab:nbbb} shows the number of signal and background
events remaining after the selections. The statistical significance
of the signal is shown in Tables~\ref{tab:sbll10}, \ref{tab:sbll12}
and \ref{tab:sbll14} for various choices of the cut on missing $E_T$.

\vspace{1cm}
  \begin{Tabhere}
  \[
  \begin{array}{|c|c||c|c|}
  \hline
  \etmiss > & \mbox{Type} & \mbox{Low luminosity},30 fb^{-1} & \mbox{High luminosity}, 100 fb^{-1} \\
  \hline
  \hline
  1\mbox{ TeV} & jZ(\nu\nu)   & 120.6 & 414.0 \\
               & jW(\tau\nu)  &  34.5 & 122.7 \\
               & jW(e\nu)     &   2.7 &   8.8 \\
               & jW(\mu\nu)   &   3.3 &  11.0 \\
               & \mbox{Total} & 161.1 & 556.5 \\
  \hline
  1.2\mbox{ TeV} & jZ(\nu\nu) &  36.1 & 124.7 \\
               & jW(\tau\nu)  &   9.2 &  30.1 \\
               & jW(e\nu)     &   0.6 &   2.0 \\
               & jW(\mu\nu)   &   0.9 &   2.9 \\
               & \mbox{Total} &  46.9 & 159.7 \\
  \hline
  1.4\mbox{ TeV} & jZ(\nu\nu) & 11.1 &  37.4 \\
               & jW(\tau\nu)  &  2.8 &   9.6 \\
               & jW(e\nu)     &  0.1 &   0.6 \\
               & jW(\mu\nu)   &  0.2 &   0.8 \\
               & \mbox{Total} & 14.3 &  48.4 \\
  \hline
  \end{array}
  \]
  \caption{Number of remaining background events after the 
  selection. Three different cuts on $\etmiss$ are shown.}
  \label{tab:nbbb}
  \end{Tabhere}

  \begin{Tabhere}
  \[
  \begin{array}{|c|c||c|c|c||c|c|c|}
  \hline
  \delta & M_{D} & \multicolumn{3}{c||}{\mbox{Low luminosity},30 fb^{-1}} & \multicolumn{3}{c|}{\mbox{High luminosity}, 100 fb^{-1}} \\
  & & S & S/\sqrt{B} & S/\sqrt{7B} & S & S/\sqrt{B} & S/\sqrt{7B} \\
  \hline
  \hline
  2 & 4 & 1036.4 & 81.6 & 30.8 & 3542.2 & 150.2 & 56.8 \\
    & 5 &  417.0 & 32.9 & 12.4 & 1426.9 & 60.4 & 22.8 \\
    & 6 &  205.9 & 16.3 &  6.2 &  700.6 & 29.6 & 11.2 \\
    & 7 &  111.3 &  8.8 &  3.3 &  379.4 & 16.1 &  6.1 \\
    & 8 &   65.3 &  5.2 &  2.0 &  222.5 &  9.4 &  3.5 \\
  \hline
  3 & 4 & 641.8 & 50.6 & 19.1 & 2168.4 & 92.0 & 34.8 \\
    & 5 & 211.5 & 16.6 &  6.3 &  706.0 & 30.0 & 11.3 \\
    & 6 &  85.1 &  6.8 &  2.6 &  287.5 & 12.1 &  4.6 \\
    & 7 &  39.3 &  3.1 &  1.2 &  134.0 &  5.7 &  2.2 \\
  \hline
  4 & 4 & 436.2 & 34.3 & 13.0 & 1473.4 & 62.5 & 23.6 \\
    & 5 & 113.0 &  8.8 &  3.3 &  383.4 & 16.3 &  6.2 \\
    & 6 &  37.8 &  2.9 &  1.1 &  128.5 &  5.4 &  2.0 \\
  \hline
  \end{array}
  \]
  \caption{Number of remaining signal events after the 
  selection ($\etmiss > 1$ TeV) and statistical significance.}
  \label{tab:sbll10}
  \end{Tabhere}

  \vspace{2cm}

  \begin{Tabhere}
  \[
  \begin{array}{|c|c||c|c|c||c|c|c|}
  \hline
  \delta & M_{D} & \multicolumn{3}{c||}{\mbox{Low luminosity},30 fb^{-1}} & \multicolumn{3}{c|}{\mbox{High luminosity}, 100 fb^{-1}} \\
  & & S & S/\sqrt{B} & S/\sqrt{7B} & S & S/\sqrt{B} & S/\sqrt{7B} \\
  \hline
  \hline
  2 & 4 & 463.8 & 67.7 & 25.6 & 1548.5 & 122.5 & 46.3 \\
    & 5 & 186.2 & 27.2 & 10.3 &  622.8 & 49.4 & 18.7 \\
    & 6 &  91.5 & 13.3 &  5.0 &  306.1 & 24.2 &  9.2 \\
    & 7 &  47.3 &  6.9 &  2.6 &  163.9 & 13.0 &  4.9 \\
    & 8 &  28.5 &  4.2 &  1.6 &   96.1 &  7.6 &  2.9 \\
  \hline
  3 & 4 & 312.1 & 45.6 & 17.2 & 1050.0 & 83.1 & 31.4 \\
    & 5 & 101.3 & 14.7 &  5.6 &  345.6 & 27.4 & 10.3 \\
    & 6 &  40.9 &  5.9 &  2.2 &  139.2 & 11.1 &  4.2 \\
    & 7 &  18.6 &  2.8 &  1.0 &   62.9 &  5.0 &  1.9 \\
  \hline
  4 & 4 & 206.1 & 30.1 & 11.4 & 691.6 & 54.7 & 20.7 \\
    & 5 &  54.0 &  8.0 &  3.0 & 184.1 & 14.5 &  5.5 \\
    & 6 &  18.1 &  2.6 &  1.0 &  60.5 &  4.8 &  1.8 \\
  \hline
  \end{array}
  \]
  \caption{Number of remaining signal events after the 
  selection ($\etmiss > 1.2$ TeV) and statistical significance.}
  \label{tab:sbll12}
  \end{Tabhere}

  \begin{Tabhere}
  \[
  \begin{array}{|c|c||c|c|c||c|c|c|}
  \hline
  \delta & M_{D} & \multicolumn{3}{c||}{\mbox{Low luminosity},30 fb^{-1}} & \multicolumn{3}{c|}{\mbox{High luminosity}, 100 fb^{-1}} \\
  & & S & S/\sqrt{B} & S/\sqrt{7B} & S & S/\sqrt{B} & S/\sqrt{7B} \\
  \hline
  \hline
  2 & 4 & 187.6 & 49.5 & 18.7 & 645.4 & 92.8 & 35.1 \\
    & 5 &  77.6 & 20.4 &  7.7 & 272.8 & 39.1 & 14.8 \\
    & 6 &  38.7 & 10.2 &  3.9 & 128.8 & 18.5 &  7.0 \\
    & 7 &  19.7 &  5.2 &  2.0 &  66.5 &  9.5 &  3.6 \\
    & 8 &  11.6 &  3.1 &  1.2 &  39.4 &  5.7 &  2.2 \\
  \hline
  3 & 4 & 142.5 & 37.8 & 14.3 & 479.8 & 68.9 & 26.1 \\
    & 5 &  46.2 & 12.3 &  4.6 & 159.8 & 23.0 &  8.7 \\
    & 6 &  18.8 &  5.0 &  1.9 &  64.0 &  9.2 &  3.5 \\
    & 7 &   8.5 &  2.3 &  0.9 &  29.4 &  4.2 &  1.6 \\
  \hline
  4 & 4 & 97.1 & 25.6 & 9.7 & 324.4 & 46.6 & 17.6 \\
    & 5 & 25.2 &  6.6 & 2.5 &  86.7 & 12.5 &  4.7 \\
    & 6 &  8.6 &  2.3 & 0.9 &  28.4 &  4.2 &  1.6 \\
  \hline
  \end{array}
  \]
  \caption{Number of remaining signal events after the 
  selection ($\etmiss > 1.4$ TeV) and statistical significance.}
  \label{tab:sbll14}
  \end{Tabhere}
  \vspace{5mm}

Table~\ref{tab:summary} shows the maximum value of $M_D$ and the
corresponding size of the extra dimensions for various
choices of $\delta$ to which ATLAS is sensitive. Values of $M_D$ below 
the values shown will be discovered with at least  $5\sigma$
significance. We have used the conservative statistical estimator in
determining these values. The table also shows
the value of $M_D$ below which the event rates quoted here are not
reliable and will be modified by the new physics appearing at scale
$M_D$. Note that for $\delta=4$ the two values of $M_D$ are quite
close; for larger values of $\delta$ the LHC sensitivity cannot be
assessed using the signal that we have described and the limit of
sensitivity is therefore model dependent.

  \begin{Tabhere}
  \[
  \begin{array}{|c|c|c|c|}
  \hline
  \delta & M_{D}^{min} \mbox{ (TeV)}& M_{D}^{max} \mbox{ (TeV)}& R_{compact}\\
  \hline
  \hline
  2 & \sim 4   & 7.5 & 10\;\mu\mbox{m} \\
  3 & \sim 4.5 & 5.9 & 300\;\mbox{pm} \\
  4 & \sim 5   & 5.3 & 1\;\mbox{pm} \\
  \hline
  \end{array}
  \]
  \caption{$M_{D}$ ranges for which a 5-sigma discovery is possible after one year at 
  high luminosity. $R$ is the equivalent radius of compactification of 
   $M_{D}^{max}$. The missing transverse energy was required to be larger 
than one TeV and $S/\sqrt{7B}$ was used as a statistical estimator.}
  \label{tab:summary}
  \end{Tabhere}

\section{Single photon signature}

Another interesting signal at LHC is the production of the graviton in association 
with a photon. However the rates are much lower than in the graviton plus jet 
cases and the region of $(\delta,M_{D})$ which can be probed is much more 
limited. We include a discussion here as this  signature could
 be used as a confirmation after a discovery in the jet 
channel. 

 \subsection{Effective theory and validity range}

\begin{figure}
  \[\begin{array}{cc}
  \mbox{\epsfig{file=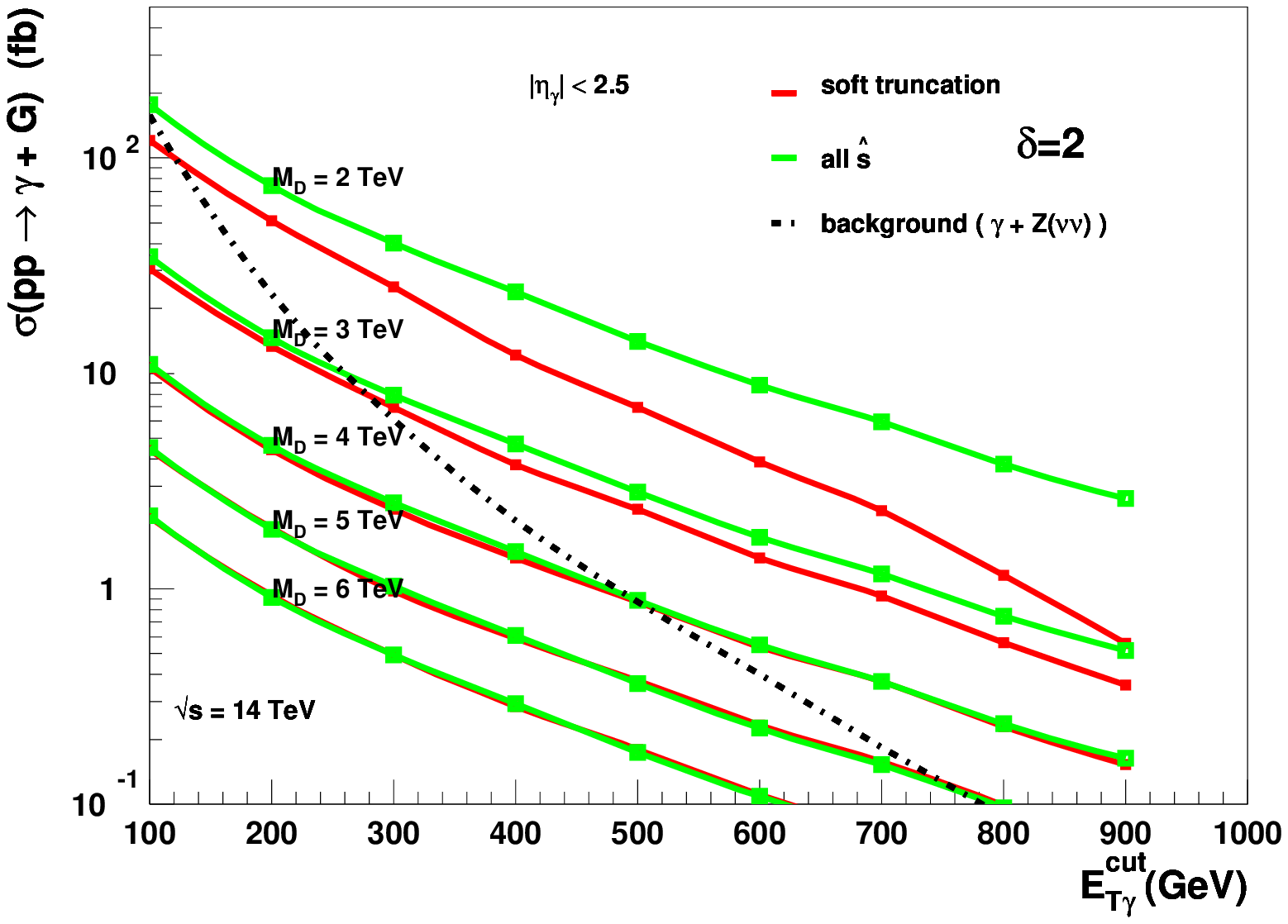,width=0.5\linewidth}}&
  \mbox{\epsfig{file=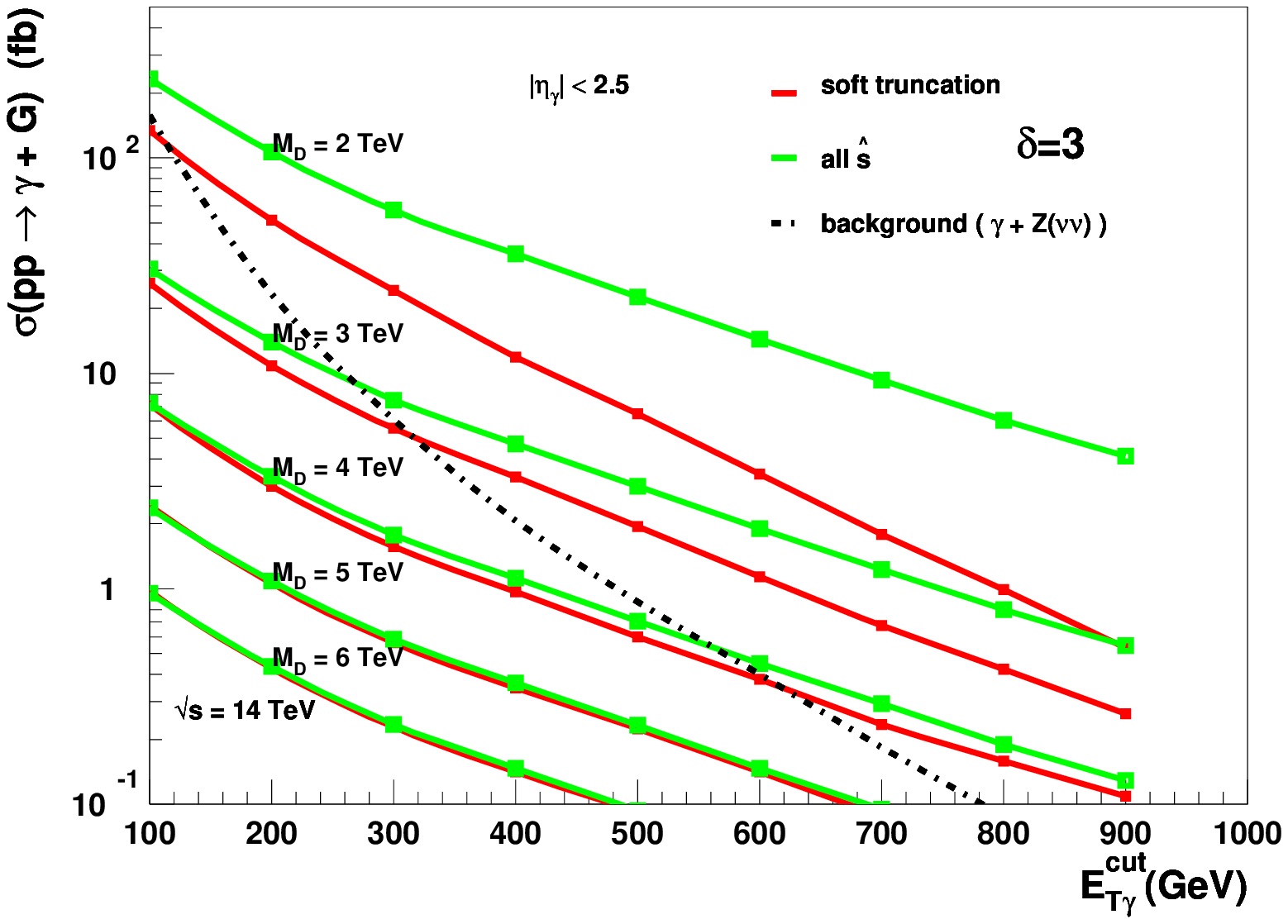,width=0.5\linewidth}}\\
  \end{array}\]
  \caption{Integrated cross-section for the graviton plus photon 
  signal as a function of the cut on the photon transverse energy 
  for $\delta=2$ (left) and $\delta=3$ (right). The dashed line shows 
  the main standard model background. The solid black lines show the 
  effect of truncating the process in ISAJET.}
  \label{fig:g-svset}
  \end{figure}

  To determine the validity range of the theory, we used the same 
  method as described in section~\ref{sec:range}. 
  Fig.~\ref{fig:g-svset} shows the cross-section 
  $\sigma(pp\rightarrow\gamma G)$ as a function of the cut on the 
  photon transverse energy ($\sim\etmiss$), for two extra-dimensions. 
  The photon
  is required to be central ($|\eta| < 2.5$). The contribution of 
  the main background ($\gamma+Z(\rightarrow\nu\nu)$) is also 
  shown on the figure. The large 
  differences between the standard cross-section and the 
  truncated one ($\hat{s}<M_{D}^2$) for $M_{D}$ below 4 TeV 
  clearly show that the
  predictions are model dependent in this range.
  Unfortunately, the rates for $M_{D} \geq 4$ TeV are very low.

  For three extra-dimensions, there is no region where the model
  independent predictions can be made and where  
  the rate is high enough to observe signal events 
  over the background (Fig.~\ref{fig:g-svset}). 
  The situation is 
  of course even worse for larger values of $\delta$.


 \subsection{Backgrounds}

 The main Standard model background is $\gamma Z$ where the $Z$ decays into 
 two neutrinos. PYTHIA 5.720~\cite{pythia} has been used to generate 
 300 000 of these events.
  As in the jet case, the second most important background is the channel 
 where a $W$ boson is produced with the photon and where a large amount of 
 missing energy can appear because of the $W$ decay to $\tau\nu$.
 A sample of 300 000 events were generated with ISAJET.

 Since the number of events with the other leptonic decays of the $W$ 
 ($e\nu$ and $\mu\nu$) which survived the lepton veto has been found to 
 be negligible for the jet channel, those backgrounds have not been 
 simulated for the photon channel.
 
 \subsection{Analysis}

For triggering, at least one photon with $E_{T}>60$ GeV at high luminosity
and within $|\eta|<2.5$ is required as well as at least 100 GeV of missing 
transverse energy.
In addition, 
the lepton veto described in section~\ref{sec:lepveto} has been applied. 
As before the signal emerges from the background  
at large  transverse missing energy: typically 
$\etmiss > 500$ GeV is required.

 \subsection{Sensitivity}

 To derive the sensitivity, a calibration sample of $\gamma Z(ee)$ events 
 has been generated and selected as follows:
 in addition to a photon fulfilling the same trigger requirements as above, 
 two electrons within $|\eta|<2.5$ and with $p_{T}>20$ GeV are required. The 
 invariant mass of the two electrons has to lie within $m_{Z}\pm 10$ GeV.
The calibration sample is 6 times smaller than the background sample. 
 Therefore the statistical estimator used is  $S/\sqrt{6B}$.
 The number of remaining background events after requiring that 
 $\etmiss>500$ GeV is summarized in table~\ref{tab:nb-g} for an 
 integrated luminosity of $100$ fb$^{-1}$. The number of signal events and 
 the corresponding sensitivities are shown in Table~\ref{tab:ns-g}.

 \vspace{5mm}
  \begin{Tabhere}
  \[
  \begin{array}{|c|c||c|c|}
  \hline
  \etmiss > & \mbox{Type} & \mbox{High luminosity}, 100 fb^{-1} \\
  \hline
  \hline
  500\mbox{ GeV} & \gamma Z(\nu\nu)   &  80.7 \\
                 & \gamma W(\tau\nu)  &   2.2 \\
                 & \mbox{Total}       &  82.9 \\
  \hline
  \end{array}
  \]
  \caption{Number of remaining background events after the selection.}
  \label{tab:nb-g}
  \end{Tabhere}
 \vspace{5mm}
  \begin{Tabhere}
  \[
  \begin{array}{|c|c||c|c|c|}
  \hline
  \delta & M_{D} & \multicolumn{3}{c|}{\mbox{High luminosity}, 100 fb^{-1}} \\
  & & S & S/\sqrt{B} & S/\sqrt{6B} \\
  \hline
  \hline
  2 & 3 & 194.4 & 21.4 & 8.7 \\
    & 4 &  61.8 &  6.8 & 2.8 \\
  \hline
  3 & 4 &  49.2 &  5.4 & 2.2 \\
  \hline
  \end{array}
  \]
  \caption{Number of remaining signal events after the selection and 
  corresponding sensitivities. Various 
  values of $\delta$ and $M_{D}$ are shown.}
  \label{tab:ns-g}
  \end{Tabhere}
  \vspace{5mm}

  As expected the sensitivity is very limited with this channel. Nevertheless 
  it could provide a valuable check in case of a discovery in the jet channel, 
  provided both the number of extra dimensions and the mass scale are not too 
  large.

 \section{Summary and interpretation}

 Using the most conservative statistical estimator, the maximum value of the 
 mass scale for which a 5-sigma discovery is possible has been derived, as 
 well as the corresponding radius of the compactified space (assumed to be 
 a torus). The results are summarized in Tables~\ref{tab:jet-mmax} 
 and~\ref{tab:gamma-mmax} for the jet and the gamma channels respectively.

 \begin{Tabhere}
  \[
  \begin{array}{|c|c|c|c|}
  \hline
  \delta & M_{D}^{min} \mbox{ (TeV)}& M_{D}^{max} \mbox{ (TeV)}& R_{compact}\\
  \hline
  \hline
  2 & \sim 4 & 7.5 & 10\;\mu\mbox{m} \\
  3 & \sim 4.5 & 5.9 & 300\;\mbox{pm} \\
  4 & \sim 5 & 5.3 & 1\;\mbox{pm} \\
  \hline
  \end{array}
  \]
 \caption{Mass scale ranges which can be probed at LHC for different 
 number of extra dimensions using the jet plus missing transverse 
 energy signature. The maximum value $M_{D}^{max}$ corresponds 
 to the highest mass scale for which $S>5\sqrt{7B}$. $R_{compact}$ is the 
 corresponding radius of compactification. $M_{D}^{min}$ is an indicative 
 value of the limit of reliability for the effective theory.}
 \label{tab:jet-mmax}
 \end{Tabhere}
 \vspace{5mm}

 \begin{Tabhere}
  \[
  \begin{array}{|c|c|c|c|}
  \hline
  \delta & M_{D}^{min} \mbox{ (TeV)}& M_{D}^{max} \mbox{ (TeV)}& R_{compact}\\
  \hline
  \hline
  2 & \sim 3.5 & 3.7 & 30\;\mu\mbox{m} \\
  \hline
  \end{array}
  \]
 \caption{Mass scale ranges which can be probed at LHC 
 using the photon plus missing transverse 
 energy signature. For $\delta>2$ there is no value of $M_{D}$ for which a 
 model independent prediction is possible and the rate is high enough.
 The maximum value $M_{D}^{max}$ corresponds
 to the highest mass scale for which $S>5\sqrt{6B}$. $R_{compact}$ is the 
 corresponding radius of compactification. $M_{D}^{min}$ is an indicative 
 value of the limit of reliability for the effective theory.}
 \label{tab:gamma-mmax}
 \end{Tabhere}
 \vspace{5mm}

 The quoted values for $M_{D}^{min}$ are indicative of the limit of 
 reliability for the effective theory: below $M_{D}$, some new 
 physics may well occur. However it should be stressed again that a signal
 is not precluded in this region. Furthermore, it is possible to extend the 
 interesting region to lower values of $M_{D}$ by releasing the cut on the 
 transverse missing energy (provided the background still remains under 
 control).
 
\label{sec:underlying-theory}

If a signal is observed at LHC, one would like to measure the
fundamental parameters $M_D$ and $\delta$. As can be seen from
Fig.~\ref{fig:summary}, using the shapes of these curves it is
not possible to determine both $\delta$ and $M_D$; for example the
curves for $\delta=2$ and $M_D=6$ TeV are very similar to those
for$\delta=3$ and $M_D=5$ TeV.

 In order to distinguish these, we must exploit the variation of the cross-section with the
center of mass energy of the LHC. Fig.~\ref{fig:svset-10} 
shows the rates at a center of mass energy of 10
TeV. It can be seen that the rates for
$\delta=2$ and $M_D=6$ TeV and $\delta=3$ and $M_D=5$ TeV differ by
approximately a factor of two.
 
There is a kinematic limit ($m^{max}$) on the mass of the graviton
that can be emitted. This limit depends on the 
the center of mass  energy.
  of respectively 14 and 10 TeV.
  (cf. Eqn.~\ref{eq:xsection}). The ratio of cross-sections at
  different center of mass energies
$$  
     \frac{ \sigma(pp\rightarrow jet+G)_{\sqrt{s}=10\;TeV} } 
          { \sigma(pp\rightarrow jet+G)_{\sqrt{s}=14\;TeV} }
$$
is almost independent of $M_D$ and varies  with $\delta$ as
can be seen from  Fig.~\ref{fig:svsmd-10-14}.
  The measured ratio  should allow  determination of the number of extra-dimensions, provided 
  that the error is small enough. 
  An experimental accuracy of the
  order of 5\% on the ratio is needed to discriminate between
  $\delta=2$ and $\delta=3$. For $M_D=6$ TeV and $\delta=2$, the
  event rate at 10 TeV is approximately 1/3 of that at 14 TeV
 and, since the event rates are small, comparable 
   integrated luminosity would be needed to
  measure the rate with sufficient statistical accuracy. Note that  the
  relative luminosity of the LHC at the two energies would also need
  to be measured with  comparable accuracy. 

  \begin{Fighere}
  \mbox{\epsfig{file=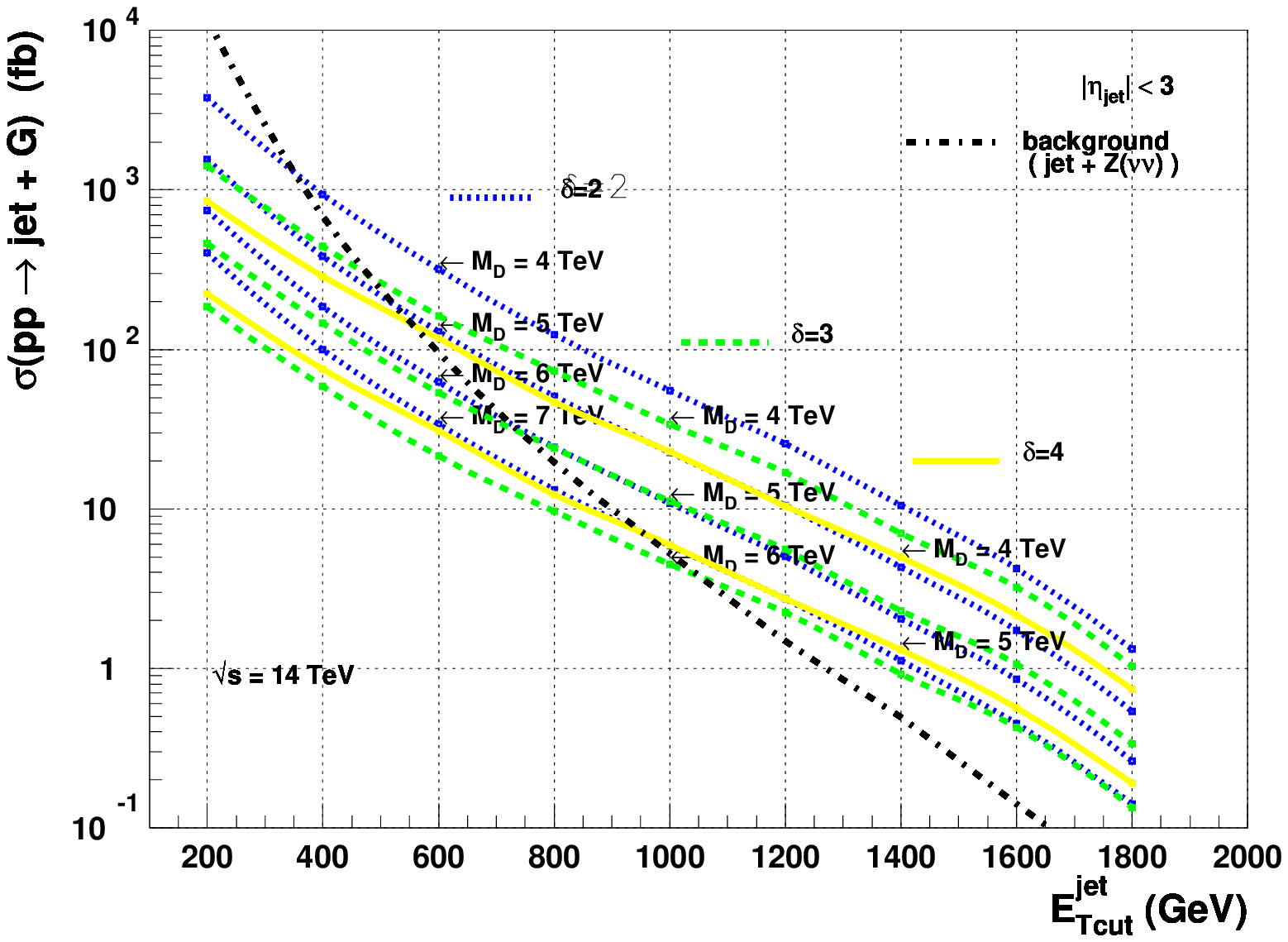,height=14cm}}
   \caption{The integrated cross-section 
   $\int{E_T>E_T^{cut_{Tjet}}}$, 
   for the
    extra dimensions processes leading to the production of a jet of
    transverse energy $E_T$ in association with missing transverse
    energy at LHC energy for various values of $\delta=2$ 
and  $M_D$. The dashed line shows the
    Standard model background.}
\label{fig:summary}
  \end{Fighere}
  \vspace{1cm}

 To estimate the sensitivity for $\sqrt{s}=10$ TeV, 
the same analysis as the one for the nominal LHC 
  center-of-mass energy (page~\pageref{sec:analysis}) has been 
  performed. 
  However
  only the 1 TeV cut on the transverse missing energy has been used. 
  We assumed the same trigger and selection criteria and
  an integrated luminosity of $50 fb^{-1}$.

 It can be seen from Fig.~\ref{fig:svset-10} that the available region 
  in the $(\delta,M_{D})$ parameter space is more limited than for a 
  center-of-mass energy of 14 TeV. It is in particular not possible to look 
  at the $\delta=4$ case.
 Fig.~\ref{fig:s10-etm} shows the distributions of the transverse 
  missing energy as a function of the cut on the jet transverse energy.

  \begin{Fighere}
  \[\begin{array}{cc}
  \mbox{\epsfig{file=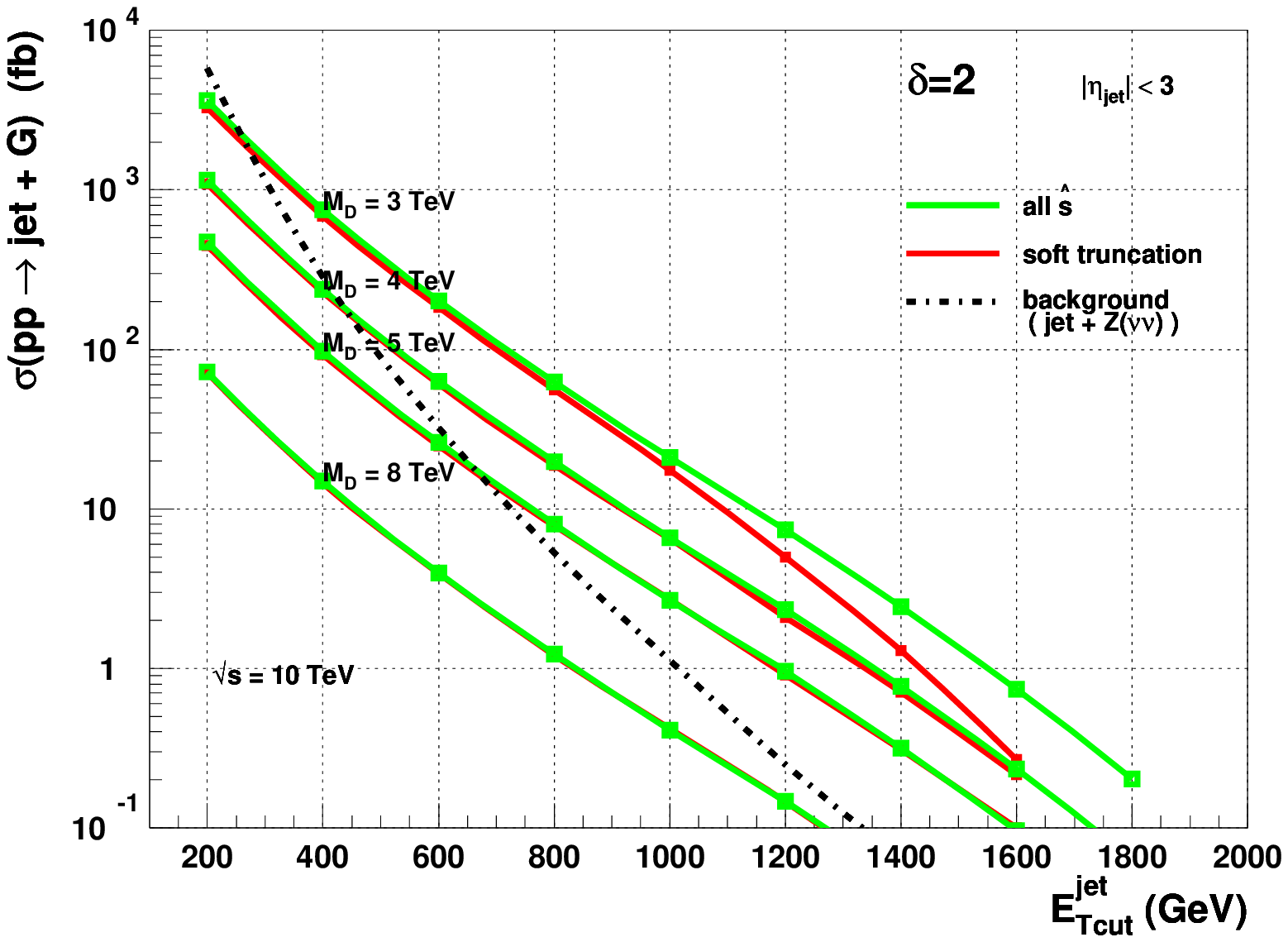,width=0.5\linewidth}}&
  \mbox{\epsfig{file=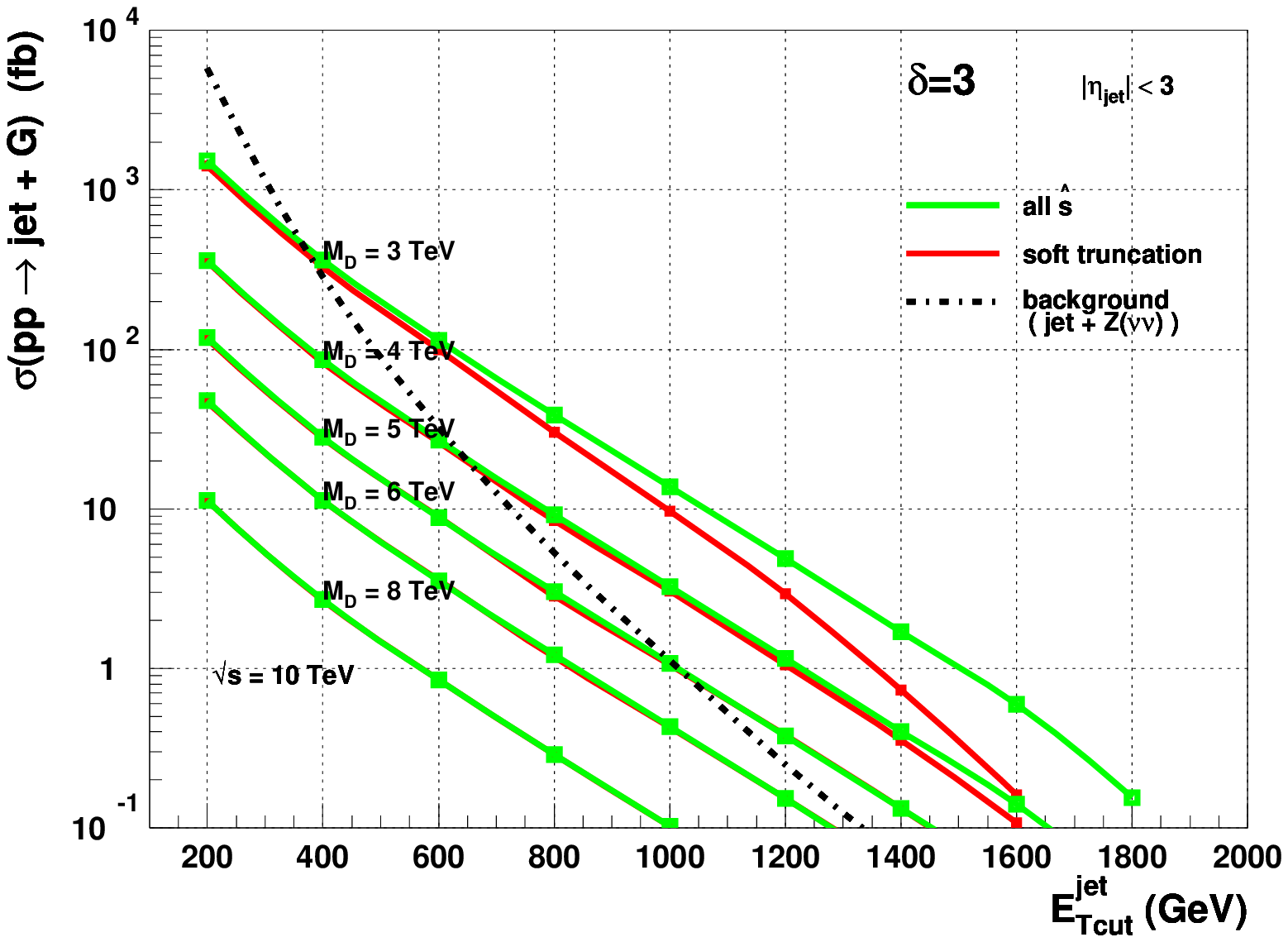,width=0.5\linewidth}}\\
  \end{array}\]
  \caption{Integrated cross-section for the graviton plus jet 
  signal as a function of the cut on the jet transverse energy 
  for $\delta=2$ (left) and $\delta=3$ (right).
  The center of mass energy is 10 TeV.}
  \label{fig:svset-10}
  \end{Fighere}
  \vspace{-1cm}
  \begin{Fighere}
  \mbox{\epsfig{file=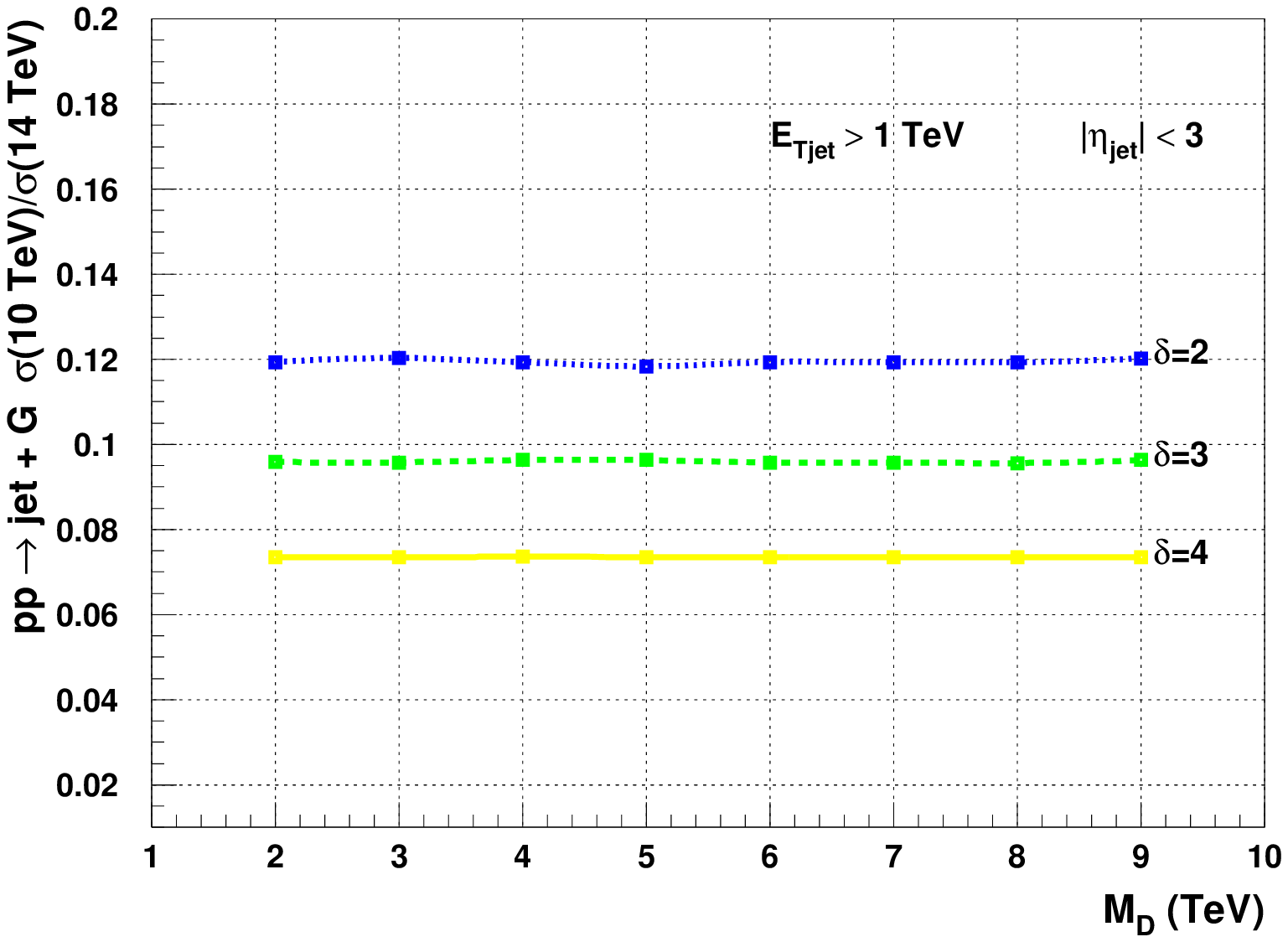,width=0.9\linewidth}}
  \caption{Distribution of the ratio of the cross-sections for the 
  $pp\rightarrow jet+G$ signal at $\sqrt{s}=10$ TeV 
  and at $\sqrt{s}=14$ TeV as a function of $M_{D}$ and for different values 
  of $\delta$.}
  \label{fig:svsmd-10-14}
  \end{Fighere}

  \begin{Fighere}
  \[\begin{array}{cc}
  \mbox{\epsfig{file=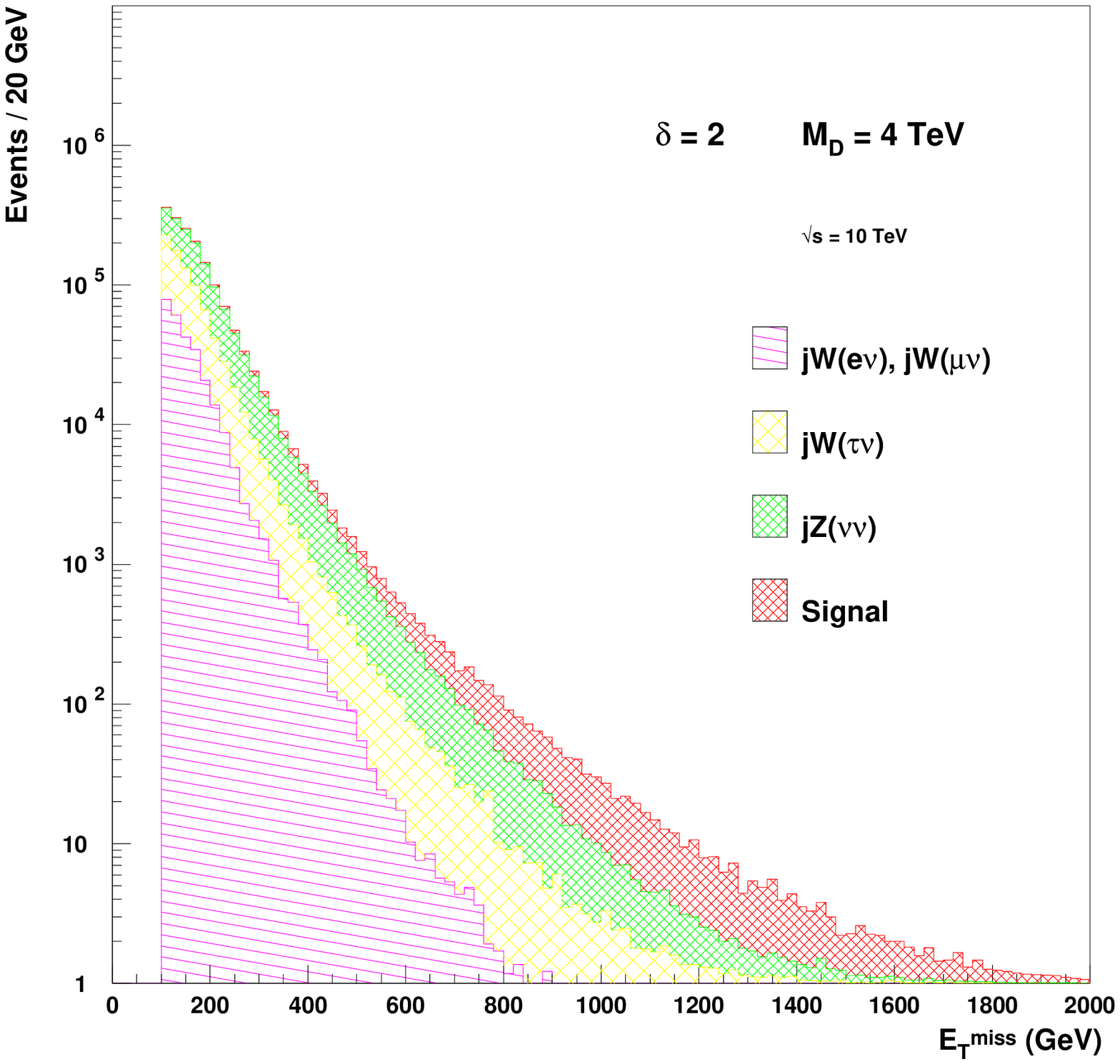,width=0.5\linewidth}}&
  \mbox{\epsfig{file=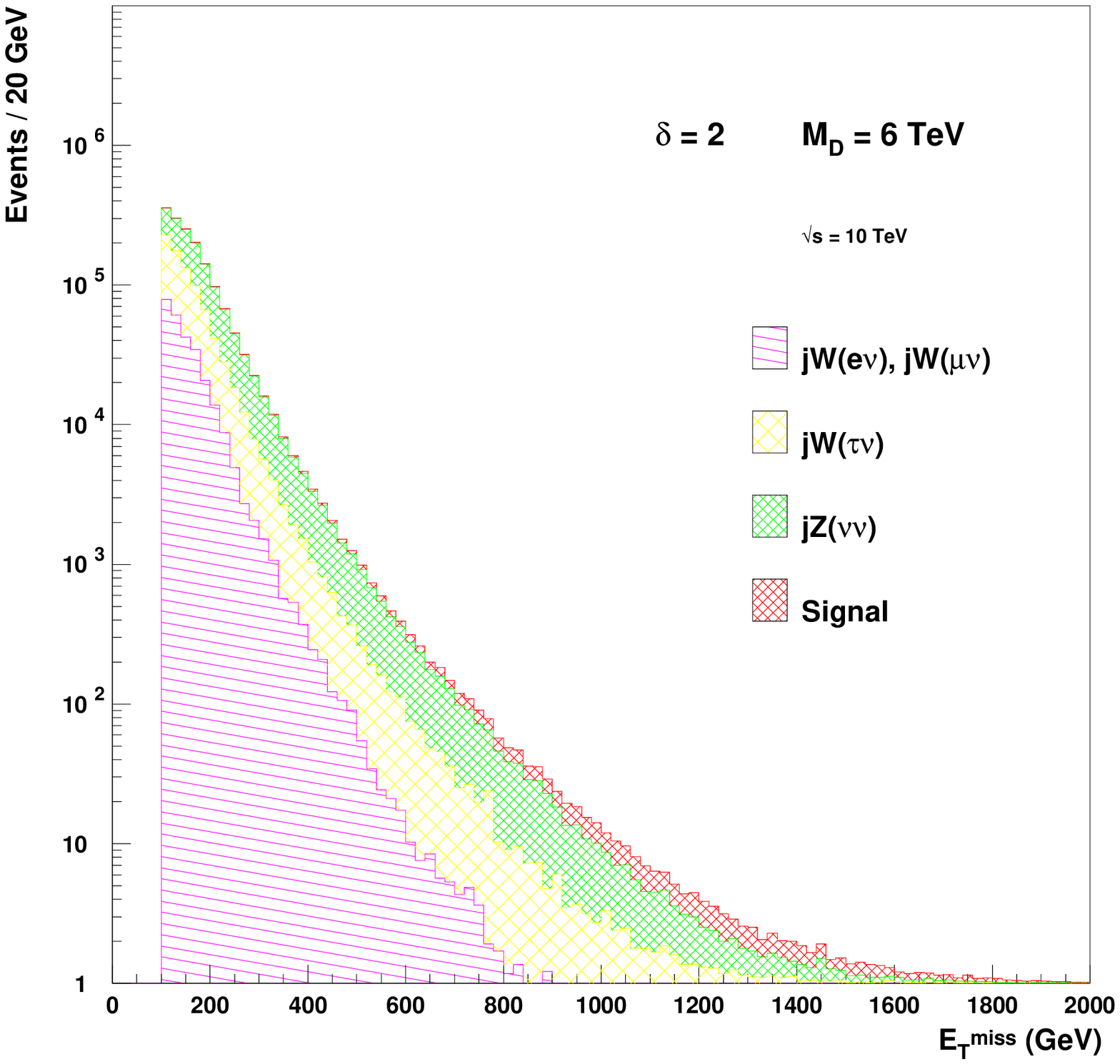,width=0.5\linewidth}}\\
  \mbox{\epsfig{file=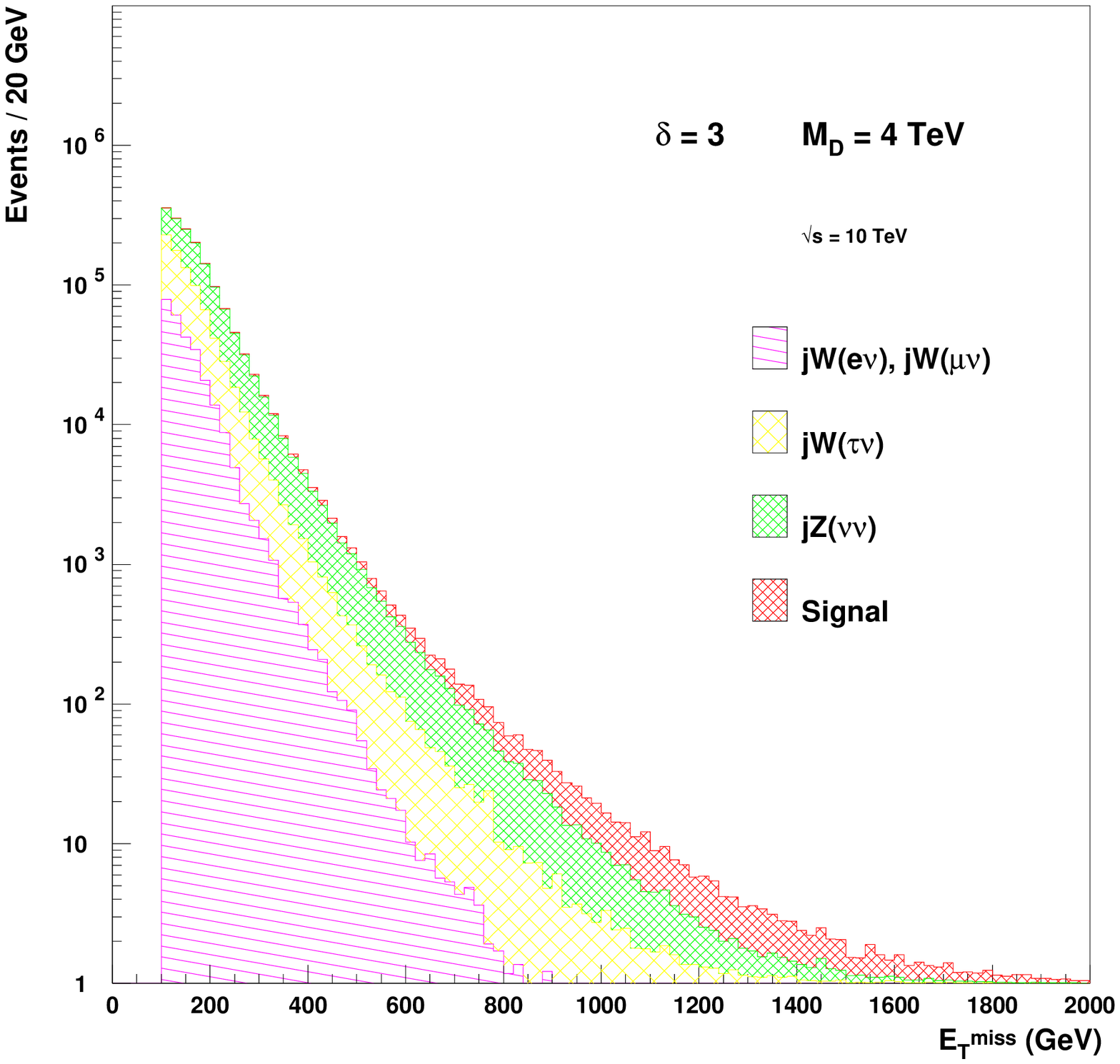,width=0.5\linewidth}}&
  \mbox{\epsfig{file=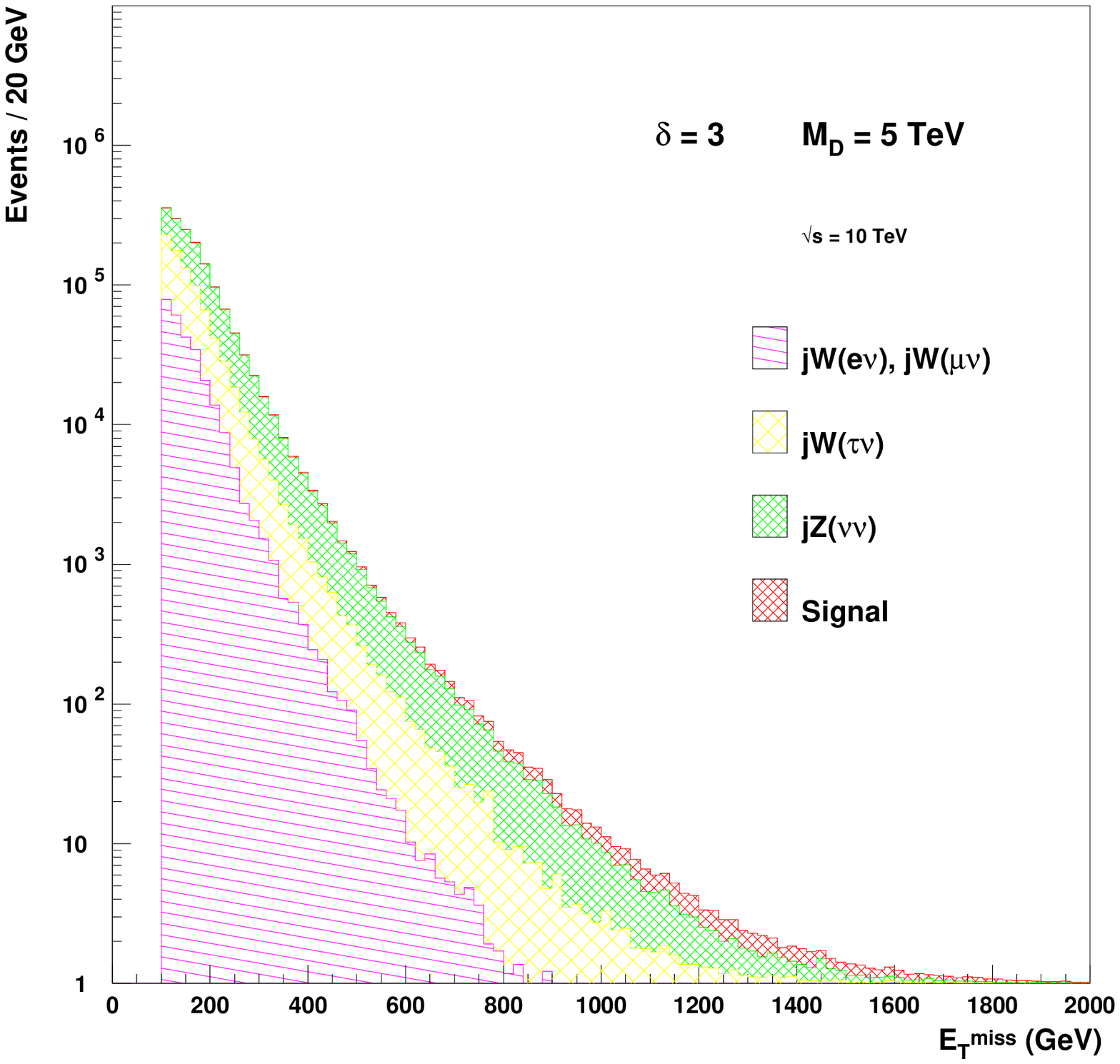,width=0.5\linewidth}}\\
  \end{array}\]
  \caption{Distributions of the missing transverse energy 
  in signal and in background events 
  after the selection and for half a year at high luminosity and at 
  $\sqrt{s}=10$ TeV. Various 
  cases $(\delta,M_{D}$) for the signal are shown.}
  \label{fig:s10-etm}
  \end{Fighere}
  \vspace{5mm}

  After the selection, the number of remaining background events and 
  signal events are shown in table~\ref{tab:s10-nbbb} and 
  table~\ref{tab:s10-sbll10} respectively. 
 The sensitivity is derived using the same method as for the 
  $\sqrt{s}=14$ TeV case. However the normalization factor is a 
  bit smaller (6) as the acceptance for $Z$ decays to leptons is
  slightly larger. Note that the method is limited to $\delta\ltap 3$
  as the rates at $10$ TeV are smaller.

  \begin{Tabhere}
  \[
  \begin{array}{|c|c||c|c|}
  \hline
  \etmiss > & \mbox{Type} & \mbox{High luminosity}, 50 fb^{-1} \\
  \hline
  \hline
  1\mbox{ TeV} & jZ(\nu\nu)   &  41.4 \\
               & jW(\tau\nu)  &  11.9 \\
               & jW(e\nu)     &   1.5 \\
               & jW(\mu\nu)   &   1.6 \\
               & \mbox{Total} &  56.4 \\
  \hline
  \end{array}
  \]
  \caption{Number of remaining background events after the 
  selection.}
  \label{tab:s10-nbbb}
  \end{Tabhere}
  \vspace{5mm}
  \begin{Tabhere}
  \[
  \begin{array}{|c|c||c|c|c|}
  \hline
  \delta & M_{D} & \multicolumn{3}{c|}{\mbox{High luminosity}, 50 fb^{-1}} \\
  & & S & S/\sqrt{3B} & S/\sqrt{6B} \\
  \hline
  \hline
  2 & 4 & 175.2 & 23.4 & 9.5 \\
    & 5 &  72.2 &  9.7 & 4.0 \\
    & 6 &  34.3 &  4.5 & 1.8 \\
  \hline
  3 & 4 &  86.5 & 11.6 & 4.7 \\
    & 5 &  28.6 &  3.8 & 1.6 \\
  \hline
  \end{array}
  \]
  \caption{Number of remaining signal events after the 
  selection ($\etmiss > 1$ TeV) and statistical significance.}
  \label{tab:s10-sbll10}
  \end{Tabhere}

 \section{Other signatures}

There are many other possible signals from theories of extra
dimensions. These involve interactions due to new particles of masses
of order $M_D$. The exchange of these particles between quarks and
leptons can induce changes in the jet and Drell-Yan rates at LHC
\cite{Hewett:1999sn}. These signals are qualitatively the same as
those for Compositeness (see for example, chapter 22 of the ATLAS Detector and Physics Performance
                                                Technical Design
                                                Report\cite{tdr} )

Another class of models have been proposed \cite{Randall:1999ee}. In
this class there is no large hierarchy between $R$ and $1/M_D$ and no
tower of graviton states. In this case the class of signatures
discussed in this note is absent.

\section*{Acknowledgments}

This work was supported in part by the Director, Office of Science,
 Office of High Energy and Nuclear Physics, Division of High Energy
Physics of the U.S. Department of Energy under Contracts
DE-AC03-76SF00098.  Accordingly, the U.S.
Government retains a non-exclusive, royalty-free license to publish or
reproduce the published form of this contribution, or allow others to
do so, for U.S. Government purposes.

\end{document}